%%%%%%%This requires the PHYZZX.TEX macropackage

 %%%%%%%If you do not have the msbm fonts, delete the following 4 lines
\font\mybb=msbm10 at 12pt
\def\bb#1{\hbox{\mybb#1}}
\def\Z {\bb{Z}}
\def\R {\bb{R}}
%%%%%%%%%%%%
%%%and replace with the following 2 lines (without %)
%\def\Z {Z}
%\def\R {R}
%%%%%%%%%%

\tolerance=10000
\input phyzzx

 \def\unit{\hbox to 3.3pt{\hskip1.3pt \vrule height 7pt width .4pt \hskip.7pt
\vrule height 7.85pt width .4pt \kern-2.4pt
\hrulefill \kern-3pt
\raise 4pt\hbox{\char'40}}}

\def\SS{{{\cal S}}}

\def\gmn{{g _{\mu \nu}}}
\def\bmn{{b _{\mu \nu}}}
\def\tgmn{{\tilde g _{\mu \nu}}}

\def\V{{\cal V}}

%%%%%%%%%%%%%%%%%%%%%%%%%%%%%%%%%%%%%%%%%%%%%%%%
%%%%%%%%%%%%%%%%%%%%
\REF\HT{C.M. Hull and P.K. Townsend, Nucl. Phys. {\bf B438} (1995) 109.}
\REF\PKT{P.K. Townsend, Phys. Lett. {\bf B350}  (1995) 184.}
\REF\Witten{E. Witten, Nucl. Phys. {\bf B443} (1995)  85, hep-th/9503124.}
\REF\HTE{C.M. Hull and P.K. Townsend, Nucl. Phys. {\bf B451} (1995) 525,
hep-th/9505073.}
\REF\AM{P.S. Aspinwall and D.R. Morrison, {\it U-Duality and Integral
Structures},
hep-th/9505025.}
\REF\Seno{A. Sen, {\it String-String Duality Conjecture in Six Dimensions and
Charged Strings},
hep-th/9504027.}
\REF\JHS {  J.A.  Harvey and A. Strominger,  {\it The Heterotic String is a
Soliton}, hep-th/9504047.}
\REF\duff{M.J. Duff, {Strong/Weak Coupling Duality from the Dual String},
hep-th/9501030.}
\REF\Oopen{C.M.Hull, Phys. Lett. {\bf B357} (1995) 545,
hep-th/9506194.}
\REF\dab{A. Dabholkar, Phys. Lett. {\bf B357} (1995) 307, hep-th/9506160.}
\REF\sdual{A. Font, L. Ibanez, D. Lust and F. Quevedo, Phys. Lett. {\bf B249}
(1990) 35; S.J. Rey, Phys. Rev. {\bf D43}  (1991) 526.}
\REF\SS {J.H. Schwarz and A. Sen, Nucl. Phys. {\bf B411} (1994) 35; Phys.
Lett. {\bf 312B} (1993) 105.}
\REF\Sen {A. Sen, Nucl. Phys. {\bf B404} (1993) 109; Phys. Lett. {\bf 303B}
(1993); Int. J. Mod. Phys. {\bf A8} (1993) 5079; Mod. Phys. Lett. {\bf
A8} (1993) 2023;    Int. J. Mod. Phys. {\bf A9} (1994) 3707.}
\REF\Cer{A. Ceresole, R. D'Auria, S. Ferrara and A. Van Proeyen,
Nucl. Phys. {\bf B444} (1995) 92, hep-th/9502072.}
\REF\Stromb{A. Strominger, Nucl. Phys. {\bf B451}, (1995) 96.}
\REF\vafa{C. Vafa, {\it A stringy test of the fate of the conifold},
hep-th/9505023.}
\REF\GMS{B.R. Greene, D.R. Morrison and A. Strominger,
 Nucl. Phys. {\bf
B451} (1995)  109 , hep-th/9504145.}
\REF\KV{S. Kachru and C. Vafa,  Nucl. Phys. {\bf B450} (1995) 69,
hep-th/9505105.}
\REF\bergort{E. Bergshoeff, C.M. Hull and T. Ortin, Nucl. Phys. {\bf B451}
(1995) 547, hep-th/9504081.}
\REF\PKTc{P.K. Townsend,   hep-th/9507048.}
\REF\asptr{P.S. Aspinwall,
hep-th/9508154.}
\REF\SSc {J.H. Schwarz, hep-th/9508143,9509148,9510086.}
\REF\Dbranes{J. Polchinski, preprint NSF-ITP-95-122, hep-th/951001.}
\REF\Dbraneswit{ E. Witten, preprint IASSNS-HEP-95-83, hep-th/9510135.}
\REF\Polwit{J. Polchinski and E. Witten, preprint IASSNS-HEP-95-81,
hep-th/9510169.}
\REF\Mwit{P. Horava and E. Witten, preprint IASSNS-HEP-95-86 hep-th/9510209.}
\REF\DGHR {A. Dabholkar, G.W. Gibbons, J.A. Harvey and F. Ruiz-Ruiz,
Nucl. Phys. {\bf B340} (1990) 33.}
\REF\HS {G.T. Horowitz and A. Strominger, Nucl. Phys. {\bf B360} (1991)
197.}
\REF\CHS {C. Callan, J. Harvey and A. Strominger, Nucl. Phys. {\bf B359}
(1991)
611.}
\REF\DL {M.J. Duff and J.X. Lu, Nucl. Phys. {\bf B354} (1991) 141;
Phys. Rev. Lett. {\bf 66} (1991) 1402;
Phys. Lett. {\bf 273B} (1991) 409.}
\REF\DLK {M.J. Duff, R.R. Khuri  and J.X. Lu,   Phys. Rep. {\bf 259} (1995)
213.}
\REF\StrFiv{
M.J. Duff, Class. Quant. Grav. {\bf 5} (1988) 189; A. Strominger, Nucl. Phys.
{\bf B343} (1990)
167; M.J. Duff and J.X. Lu, Nucl. Phys. {\bf B354} (1991) 129 and Nucl. Phys.
{\bf B357} (1991)
534. }
\REF\GrWitt{M.B. Green and E. Witten, unpublished.}
\REF\bergtown{E. Bergshoeff, M.B. Green, G. Papadopoulos and P.K. Townsend,
hep-th/9511079.}
\REF\cham{A.H.  Chamsedine, hep-th/9510100.}
\REF\CJ {E. Cremmer and B. Julia, Phys. Lett. {\bf 80B} (1978) 48; Nucl.
Phys. {\bf B159} (1979) 141.}
\REF\Shenker{S. Shenker, in {\it Random Surfaces and Quantum Gravity}, O.
Alvarez, E. Marinari
and P. Windey (eds.), Plenum (1991) 191.}
\REF\sevbrane{G.W. Gibbons, M.B.  Green and M.J. Perry, hep-th/9511080.}
\REF\DS {M.J. Duff and K.S. Stelle, Phys. Lett. {\bf 253B} (1991) 113.}
\REF\Gu {R. G{\" u}ven, Phys. Lett. {\bf 276B} (1992) 49.}
\REF\PKTb{P.K. Townsend,  Phys. Lett. {\bf B354},  (1995)  247,
hep-th/9504095.}
\REF\AMK{P.S. Aspinwall and D.R. Morrison, {\it String Theory on $K3$
Surfaces}, hep-th/9404151,
to appear in \lq Essays on Mirror Manifolds 2'.}

%%%%%%%%%%%%%%%%%%%%

%%%%%%%%%%%%%%%%%%%%%%%%%%%%%%%%%%%%%%%%%%%%%%%%%%%%%%%%%%%%%%%%%%%%
%%%%%%%%%%%%%%%%%%%%%%%%%%%%%%%%%%%%%%%%%%%%%%%%%%%%%%%%%%%%%%%%%%%%

\Pubnum{ \vbox{ \hbox {QMW-95-50}  \hbox{hep-th/9512181}} }
\pubtype{}
\date{ December 1995}

\titlepage

\title {\bf  STRING DYNAMICS AT STRONG COUPLING }

\author{C.M. Hull}
\address{Physics Department,
Queen Mary and Westfield College,
\break
Mile End Road, London E1 4NS, U.K.}

\abstract{
The dynamics of superstring,  supergravity   and M theories and their
compactifications are
probed by studying the various perturbation theories that emerge   in the
strong and weak coupling limits for various directions in coupling constant
space.
The results support the picture of an underlying non-perturbative theory  that,
when expanded
perturbatively in different coupling constants, gives different perturbation
theories, which can be perturbative superstring theories or superparticle
theories.
The $p$-brane spectrum is considered in detail and a criterion found to
establish which $p$-branes govern the strong coupling dynamics. In many cases
there are competing conjectures in the literature, and   this analysis decides
between them.
In other cases, new results are found.
The chiral six-dimensional theory resulting from compactifying the type IIB
string on $K_3$ is studied in detail and it is found that certain strong
coupling limits appear to give new theories, some of which hint at the
possibility of a twelve-dimensional origin.
}

\endpage
\pagenumber=1

%%%%%%%

%%%%%%%%%%%%%%%%%%%%%%%%%%%%%%%%
%
% S-Tables Macro
%
\message{S-Tables Macro v1.0, ACS, TAMU (RANHELP@VENUS.TAMU.EDU)}
%
% Help Text
%
\newhelp\stablestylehelp{You must choose a style between 0 and 3.}%
\newhelp\stablelinehelp{You should not use special hrules when stretching
a table.}%
\newhelp\stablesmultiplehelp{You have tried to place an S-Table inside another
S-Table.  I would recommend not going on.}%
%
% Line Thicknesses (Values)
%
\newdimen\stablesthinline
\stablesthinline=0.4pt
\newdimen\stablesthickline
\stablesthickline=1pt
%
% Border and Internal Line Thicknesses
%
\newif\ifstablesborderthin
\stablesborderthinfalse
\newif\ifstablesinternalthin
\stablesinternalthintrue
\newif\ifstablesomit
\newif\ifstablemode
\newif\ifstablesright
\stablesrightfalse
%
% Save Registers
%
\newdimen\stablesbaselineskip
\newdimen\stableslineskip
\newdimen\stableslineskiplimit
%
% Counters
%
\newcount\stablesmode
\newcount\stableslines
\newcount\stablestemp
\stablestemp=3
\newcount\stablescount
\stablescount=0
\newcount\stableslinet
\stableslinet=0
%
% Table Style Selection
%
% 0 - Centered
% 1 - Left Justified
% 2 - Right Justified
% 3 - Not Justified
%
\newcount\stablestyle
\stablestyle=0
%
% Element Buffering Definitions
%
\def\stablesleft{\quad\hfil}%
\def\stablesright{\hfil\quad}%
%
% Vertical Bar Activation
%
\catcode`\|=\active%
%
% Strut Control
%
\newcount\stablestrutsize
\newbox\stablestrutbox
\setbox\stablestrutbox=\hbox{\vrule height10pt depth5pt width0pt}
\def\stablestrut{\relax\ifmmode%
                         \copy\stablestrutbox%
                       \else%
                         \unhcopy\stablestrutbox%
                       \fi}%
%
% Misc. Internal Stuff
%
\newdimen\stablesborderwidth
\newdimen\stablesinternalwidth
\newdimen\stablesdummy
\newcount\stablesdummyc
\newif\ifstablesin
\stablesinfalse
%
% Table Macros
%
\def\begintable{\stablestart%
  \stablemodetrue%
  \stablesadj%
  \halign%
  \stablesdef}%
\def\stablesadj{%
  \ifcase\stablestyle%
    \hbox to \hsize\bgroup\hss\vbox\bgroup%
  \or%
    \hbox to \hsize\bgroup\vbox\bgroup%
  \or%
    \hbox to \hsize\bgroup\hss\vbox\bgroup%
  \or%
    \hbox\bgroup\vbox\bgroup%
  \else%
    \errhelp=\stablestylehelp%
    \errmessage{Invalid style selected, using default}%
    \hbox to \hsize\bgroup\hss\vbox\bgroup%
  \fi}%
\def\stablesend{\egroup%
  \ifcase\stablestyle%
    \hss\egroup%
  \or%
    \hss\egroup%
  \or%
    \egroup%
  \or%
    \egroup%
  \else%
    \hss\egroup%
  \fi}%
\def\stablestart{%
  \ifstablesin%
    \errhelp=\stablesmultiplehelp%
    \errmessage{An S-Table cannot be placed within an S-Table!}%
  \fi
  \global\stablesintrue%
  \global\advance\stablescount by 1%
  \message{<S-Tables Generating Table \number\stablescount}%
  \begingroup%
  \stablestrutsize=\ht\stablestrutbox%
  \advance\stablestrutsize by \dp\stablestrutbox%
  \ifstablesborderthin%
    \stablesborderwidth=\stablesthinline%
  \else%
    \stablesborderwidth=\stablesthickline%
  \fi%
  \ifstablesinternalthin%
    \stablesinternalwidth=\stablesthinline%
  \else%
    \stablesinternalwidth=\stablesthickline%
  \fi%
  \tabskip=0pt%
  \stablesbaselineskip=\baselineskip%
  \stableslineskip=\lineskip%
  \stableslineskiplimit=\lineskiplimit%
  \offinterlineskip%
  \def\borderrule{\vrule width \stablesborderwidth}%
  \def\internalrule{\vrule width \stablesinternalwidth}%
  \def\thinline{\noalign{\hrule height \stablesthinline}}%
  \def\thickline{\noalign{\hrule height \stablesthickline}}%
  \def\trule{\omit\leaders\hrule height \stablesthinline\hfill}%
  \def\ttrule{\omit\leaders\hrule height \stablesthickline\hfill}%
  \def\tttrule##1{\omit\leaders\hrule height ##1\hfill}%
  \def\stablesel{&\omit\global\stablesmode=0%
    \global\advance\stableslines by 1\borderrule\hfil\cr}%
  \def\el{\stablesel&}%
  \def\elt{\stablesel\thinline&}%
  \def\eltt{\stablesel\thickline&}%
  \def\elttt##1{\stablesel\noalign{\hrule height ##1}&}%
  \def\elspec{&\omit\hfil\borderrule\cr\omit\borderrule&%
              \ifstablemode%
              \else%
                \errhelp=\stablelinehelp%
                \errmessage{Special ruling will not display properly}%
              \fi}%
  \def\stmultispan##1{\mscount=##1 \loop\ifnum\mscount>3 \stspan\repeat}%
  \def\stspan{\span\omit \advance\mscount by -1}%
  \def\multicolumn##1{\omit\multiply\stablestemp by ##1%
     \stmultispan{\stablestemp}%
     \advance\stablesmode by ##1%
     \advance\stablesmode by -1%
     \stablestemp=3}%
  \def\multirow##1{\stablesdummyc=##1\parindent=0pt\setbox0\hbox\bgroup%
    \aftergroup\emultirow\let\temp=}
  \def\emultirow{\setbox1\vbox to\stablesdummyc\stablestrutsize%
    {\hsize\wd0\vfil\box0\vfil}%
    \ht1=\ht\stablestrutbox%
    \dp1=\dp\stablestrutbox%
    \box1}%
  \def\stpar##1{\vtop\bgroup\hsize ##1%
     \baselineskip=\stablesbaselineskip%
     \lineskip=\stableslineskip%
     \lineskiplimit=\stableslineskiplimit\bgroup\aftergroup\estpar\let\temp=}%
  \def\estpar{\vskip 6pt\egroup}%
  \def\stparrow##1##2{\stablesdummy=##2%
     \setbox0=\vtop to ##1\stablestrutsize\bgroup%
     \hsize\stablesdummy%
     \baselineskip=\stablesbaselineskip%
     \lineskip=\stableslineskip%
     \lineskiplimit=\stableslineskiplimit%
     \bgroup\vfil\aftergroup\estparrow%
     \let\temp=}%
  \def\estparrow{\vfil\egroup%
     \ht0=\ht\stablestrutbox%
     \dp0=\dp\stablestrutbox%
     \wd0=\stablesdummy%
     \box0}%
  \def|{\global\advance\stablesmode by 1&&&}%
  \def\|{\global\advance\stablesmode by 1&\omit\vrule width 0pt%
         \hfil&&}%
  \def\vt{\global\advance\stablesmode by 1&\omit\vrule width \stablesthinline%
          \hfil&&}%
  \def\vtt{\global\advance\stablesmode by 1&\omit\vrule width
\stablesthickline%
          \hfil&&}%
  \def\vttt##1{\global\advance\stablesmode by 1&\omit\vrule width ##1%
          \hfil&&}%
  \def\vtr{\global\advance\stablesmode by 1&\omit\hfil\vrule width%
           \stablesthinline&&}%
  \def\vttr{\global\advance\stablesmode by 1&\omit\hfil\vrule width%
            \stablesthickline&&}%
  \def\vtttr##1{\global\advance\stablesmode by 1&\omit\hfil\vrule width ##1&&}%
  \stableslines=0%
  \stablesomitfalse}
\def\stablesdef{\bgroup\stablestrut\borderrule##\tabskip=0pt plus 1fil%
  &\stablesleft##\stablesright%
  &##\ifstablesright\hfill\fi\internalrule\ifstablesright\else\hfill\fi%
  \tabskip 0pt&&##\hfil\tabskip=0pt plus 1fil%
  &\stablesleft##\stablesright%
  &##\ifstablesright\hfill\fi\internalrule\ifstablesright\else\hfill\fi%
  \tabskip=0pt\cr%
  \ifstablesborderthin%
    \thinline%
  \else%
    \thickline%
  \fi&%
}%
\def\endtable{\advance\stableslines by 1\advance\stablesmode by 1%
   \message{- Rows: \number\stableslines, Columns:  \number\stablesmode>}%
   \stablesel%
   \ifstablesborderthin%
     \thinline%
   \else%
     \thickline%
   \fi%
   \egroup\stablesend%
\endgroup%
\global\stablesinfalse}
%
% end of STABLES.TEX
%

\chapter{Introduction}

String theory is defined perturbatively through a set of rules for calculating
scattering amplitudes.
However, recent progress has led to some striking conjectures regarding the
non-perturbative
structure of the theory that have passed many tests [\HT-\Mwit]. The picture
that seems to be
emerging is that there is some as yet unknown theory that, when expanded
perturbatively, looks like a
perturbative string theory, but which has a surprisingly   simple structure at
the non-perturbative
level which includes U or S duality symmetries relating perturbative states to
solitons, and weak
coupling to strong. Moreover, there are a number of different coupling
constants corresponding to
 the expectation values of various scalars, and the perturbation expansions
with respect to some of
these define  string theories, but    different string theories arise for
different  coupling
constants. This leads to unexpected equivalences between string theories that
look very
different in perturbation theory: they result from different perturbation
expansions of the same
theory. In many cases, the    strong coupling limit of a given theory with
respect to
 a particular coupling constant is described by the weak coupling expansion of
a dual theory, which
is sometimes another string theory and sometimes a field theory.

An example which illustrates  many of these points is the one obtained from
the
toroidal compactification of the heterotic string to four dimensions
on $T^6$.
When the full non-perturbative theory, including solitons, is considered, there
is strong evidence
that the theory has an $SL(2, \Z)$
   S-duality symmetry relating strong to weak coupling and interchanging
electric and magnetic
charges [\sdual,\SS,\Sen]. The theory  is then self-dual: the strong coupling
limit is described by
the weak-coupling expansion of a dual heterotic string theory, which is of
exactly the same form,
but with magnetic charges   arising in the  perturbative spectrum while
electric ones arise as
solitons. Expanding the same theory in other  directions  in coupling constant
space can give the
perturbative expansion of the type IIA string or of the type IIB string
compactified on $K_3 \times
T^2$ [\HT], leading to the conjectured equivalence of the type II and heterotic
strings. The
expansion with respect to other coupling constants of the theory has been
considered in [\AM].

The $p$-brane states of the theory play a crucial role in understanding the
non-perturbative
structure [\HT]. These are associated with $p$-brane solutions of the effective
low-energy
supergravity theory that saturate a Bogomolnyi bound. As some of these
solutions are singular,
  the
question arises as to   whether they should be associated with states   in the
quantum
spectrum.
For example, type II superstring theories in ten dimensions have a string and a
5-brane coupling to
the  2-form in the NS-NS (Neveu-Schwarz/Neveu-Schwarz) sector together with
various $p$-branes
coupling to the
antisymmetric tensor gauge fields in the RR (Ramond-Ramond) sector [\HS,\DLK].
The 5-brane is
non-singular and can be regarded as a soliton of weakly-coupled string theory
[\CHS,\DL]. The
 NS-NS 1-brane solution [\DGHR] is singular, but should be regarded as the
field configuration
outside a fundamental string source. Many of the RR $p$-brane solutions are
singular, but can each be
regarded as the field configuration outside a D-brane source [\Dbranes]. The
NS-NS solitons arise as
conventional conformally invariant sigma-models, while the RR branes arise as
D-branes [\Dbranes].
 Alternatively, the singular
$p$-brane solutions of the IIA theory have a non-singular origin in the
$11$-dimensional theory,
so that the $11$-dimensional picture allows all $p$-branes to be included in
the theory [\HTE].

However, although an attractive picture is emerging, there remain many open
questions.
For example,  the strong coupling limit of the $SO(32)$ heterotic string in $
10$ dimensions  has
been variously conjectured to be a five-brane theory [\StrFiv,\DL] or a type I
superstring theory
[\Witten,\GrWitt] or
$11$-dimensional supergravity compactified on a one-dimensional degeneration of
$K_3$ [\AM], and
the question arises as to which, if any, of these is correct.
The first two appear at first to be on a very similar footing. A change of
variables in the heterotic
string low-energy effective action gives the      low-energy effective action
of the type I string
[\Witten] while a different change of variables gives a supergravity theory
with a 6-form gauge
potential (instead of a 2-form potential) that has been proposed as the
low-energy effective action
of some as yet unknown five-brane theory [\StrFiv,\DL], which might bear a
similar relation to the
6-form version of the  supergravity theory as M-theory does to 11-dimensional
supergravity.
  In each case, the change of variables includes a change in the sign of the
dilaton, so that the strong coupling limit of the heterotic string could agree
with the weak
coupling limit of either the type I string or five-brane. In
[\Oopen,\dab,\Polwit] it was argued
that the heterotic string arises as a soliton of the type I theory, but it is
straightforward to see
that it also emerges as a soliton of the proposed five-brane effective action.
String/five-brane
duality in
$D=10$  is supported by its relation
to
string/string duality in $D=6$ [\duff] (further evidence for which is proposed
in  [\Seno,\JHS]),
while the heterotic/type I duality is supported by its relation to the
$SL(2,\Z)$ duality of the
type IIB string [\Oopen], and the   evidence presented in [\Polwit].
So how are we to choose between these two conjectures, and
how are  they related to that of [\AM]?

Another issue is that of the strong coupling limit of the type IIA theory.
Similar arguments to those used in the heterotic string suggest that this might
be a type II
five-brane theory, while those of [\PKT,\Witten] suggest an $11$-dimensional
supergravity,
supermembrane or M theory.   In [\Witten] it was argued that the $0$-brane
solitons or extreme black
holes of the type IIA theory in $D=10$ have masses that scale as $g^{-1}$ where
$g$ is the string
coupling constant. These become massless in the strong coupling limit and can
be identified with
the Kaluza-Klein modes of $11$-dimensional supergravity compactified on a
circle of radius $R \sim
g^{2/3}$ [\Witten], leading to the conjecture that the strong coupling limit is
$11$-dimensional
supergravity.
These \lq extreme black holes' are in fact singular solutions of the $D=10$
supergravity theory,
but can be associated with $p=0$ D-branes.
Moreover, the theory also has RR $p$-branes with $p=2,4,6$ (and 8
[\Dbranes,\bergtown]) whose mass per unit $p$-volume
also scales as $g^{-1}$, together with  a NS-NS five-brane whose density scales
as $g^{-2}$  so that
these all appear to become \lq massless' in the strong coupling limit. More
properly, they appear
to  become null $p$-branes with vanishing density and zero tension whose
world-volume is a null
$(p+1)$-surface. These extra massless solitons could spoil the interpretation
as an
$11$-dimensional field theory; indeed, the fact that the five-brane appears to
become massless
faster than the other $p$-branes might be taken as evidence in favour of a dual
five-brane   theory.
Even if the 11-dimensional conjecture is accepted, there remains the question
as to whether the
strong coupling limit is a supergravity theory as suggested in [\Witten] or
whether it is some
supermembrane or M theory, since the analysis of [\Witten] only addresses a
particular class of
particle states.

Similar remarks apply to the type IIB theory, which again could be dual to a
type IIB 5-brane theory.
On the other hand, the  type IIB string is conjectured to have an
$SL(2,\Z)$ U-duality symmetry [\HT], which would imply that it is self-dual:
the strong coupling
limit is again a type IIB superstring theory [\Witten] in which the RR string
of the original
theory becomes fundamental, through a ten-dimensional string-string duality
[\Oopen].
In addition to the NS-NS string which is fundamental  at weak coupling, the
theory has    a
NS-NS 5-brane whose density scales as $g^{-2}$ and RR $p$-branes for
$p=1,3,5,7,9$ whose
densities scale as $g^{-1}$ [\Dbranes]. String/5-brane duality would require
that the NS-NS 5-brane
governs the strong coupling dynamics, while string-string duality would require
that it is the RR
string  that does so. However, all the $p$-branes have densities that become
small at strong
coupling and in particular the NS-NS 5-brane
is the one whose  density (with respect to the string-metric)
 tends to zero \lq fastest'. It is important to understand the strong coupling
dynamics in this
case, and in particular to   test   the predictions of U-duality.

Conjectures for the strong coupling limits of many string theories are to be
found in [\Witten], but
a number of
  gaps remain. One of the most interesting concerns the strong coupling
behaviour of the type IIB
theory compactified on $K_3$. As will be seen, going to infinity in certain
directions in the coupling
constant space of this theory gives limits that do not seem to correspond to
any known theories and
might   arise from  a new theory in twelve dimensions.

The purpose of this paper is to address these and related issues.
In [\Witten] the strong coupling limit of various string theories was
investigated by seeking the
$0$-brane or particle states that became massless fastest as one went to
infinity in a particular
direction in coupling constant space.
Here, this will be generalised to a study of the
  $p$-brane states (which may be represented at weak coupling by solitons or
fundamental branes or
D-branes)  that become massless or null in superstring or supergravity or M
theories in certain weak
or strong coupling limits.  It does not make sense to ask which of these
$p$-branes   become massless
\lq fastest', since the masses of $p$-branes with different values of $p$
cannot be compared.\foot{
If there are some compact dimensions, one can compare the masses of particle
states in the
compactified theory arising from
$p$-branes wrapping around homology cycles, but it is clear from [\Witten] that
the  relationship
between the strong coupling dynamics of compactified theories and that of
uncompactified ones can be
subtle, and it is desirable to give an analysis that does not rely on
compactification, but which
gives the expected results if some dimensions are compactified.}
 Instead, the question as to which $p$-branes govern the weak or
strong coupling behaviour is addressed by finding a criterion to establish
which correspond to
perturbative states at weak coupling and which to perturbative states of the
strong coupling theory,
when considered as a perturbation theory in the inverse coupling. If the
perturbative states are
one-branes, then the perturbative theory is a string theory, while if they are
0-branes it is a
field theory, and so on. The analysis of [\Witten,\AM]   gave the perturbative
particle
states of the strong coupling limit of various string theories and hence
identified the supergravity
theory that described the low-energy effective dynamics at strong coupling, the
 analysis presented
here goes some way to identifying the cases in which the perturbation theory
that arises in the strong coupling limit is a superstring
theory and the cases in which it is a superparticle theory.

The key to the analysis is to study the coupling constant dependence of the
mass/$p$-volume of the
$p$-branes. For example, in $N=4$ supersymmetric Yang-Mills in $D=4$ with the
gauge group
spontaneously broken to an abelian subgroup, the masses are conventionally
defined so that particles
carrying electric charge have masses that are independent of the coupling
constant
$g$, while those carrying magnetic charge have $M \sim 1/g^2$, so that in the
weak coupling limit $g
\to 0$ the magnetic charges have infinite mass and  only the electric charges
remain.
Perturbation theory is then based on these electric charges (together with
neutral particles) and it
is these that propagate in Feynman diagrams, while the magnetic charges are
treated as solitons.
To study strong coupling, it is convenient to make certain rescalings so that
the magnetic charges
have
$M
\sim 1$ while the electric charges have $M \sim g^2 $. Then the electric
charges decouple in the
strong coupling limit $g \to \infty$ and are treated as solitons, while the
magnetic  charges
(plus neutral states) are   perturbative states for  the perturbation theory
arising from
the expansion in   $\hat g=1/g$, which is small for large $g$.
There is perhaps an underlying theory in which electric and magnetic charges
appear symmetrically,
but when treated perturbatively in $g$, the electric charges
are   perturbative states and the magnetic charges are solitonic, while
when treated perturbatively in  $\hat g$, the magnetic  charges
are   perturbative states and the  electric charges are solitonic.
The \lq perturbative states' are the fundamental objects of the usual field
theory description, and
the strong coupling field theory is again an $N=4$ super-Yang-Mills theory as
expected from the
Montonen-Olive-Osborne duality conjecture. However, the conventional field
theory picture is perhaps
an artifice of using perturbation theory and at intermediate values of $g$
field theory is not so
useful. Indeed, at enhanced symmetry points both electric and magnetic charges
become massless
[\HTE], so that a conventional local  field  theory description cannot  be
applied. However, until
the underlying theory with electric and magnetic charges treated \lq
democratically' is understood,
important progress can be made by studying the effective perturbation theory in
$g$ or $\hat g$.

As another example, consider the type IIA or IIB string in ten dimensions. In
the string metric,
the NS-NS  string has density $M_1 $ that is independent of the string coupling
$g$, while the RR
$p$-branes have density $M_p \sim 1/g$ and the NS-NS 5-brane has density $M_5
\sim 1/g^2$. This is
consistent with treating the NS-NS string as the perturbative state  in an
expansion in $g$, so
that a conventional string perturbation theory emerges, while the other
$p$-branes decouple as $g \to 0$ and should be treated  as solitons.

A similar picture extends to other  theories with local supersymmetry, such as
superstrings,
M-theory or supergravities.  The $p$-brane spectrum can be studied by
consideration of the low-energy
effective supergravity theory.  In cases with enough unbroken
supersymmetry, the densities of BPS-saturated $p$-branes, and in particular
their coupling
constant dependence,
 is reliably given by the Bogomolnyi formula.
The  coupling constant dependence can be changed by rescaling the metric etc,
but in each case there
are two particularly natural choices of metric etc. In the first, some of the
$p$-branes have
densities that are independent of a particular coupling constant $g$, and all
other densities
depend on $g$ to a {\it negative} power. This suggests treating the states with
$M_p\sim 1$ as
perturbative states in a weak coupling expansion, and the others with $M_p\sim
g^{-n}$ (for some
$n$) as solitons. With the second choice, there is some new set of states with
$M_p \sim 1$ and all others depend on $g$ to a {\it positive } power.
This suggests treating the states with $M_p\sim 1$ as
perturbative states in a strong coupling expansion in $\hat g =1/g$, and the
others with $M_p\sim
\hat g^{-n}$   as solitons. In either case, this identifies those BPS-saturated
$p$-brane states that
should be included in the relevant perturbative spectrum, together with other
states including
neutral massless ones. These are the states that should propagate in loops. In
some cases, the
perturbative spectrum consists of strings and a consistent superstring
perturbation theory
emerges, while in others it is that of a supergravity theory in which case the
corresponding
perturbation theory only makes sense in the presence of a cut-off.
We shall see that this analysis gives the expected results in many cases where
alternative methods
(e.g. using U-duality) can be used, and gives a consistent picture in other
cases.

In the cases examined here, the perturbative theory that emerges is
almost always either a supergravity theory or a superstring theory. Moreover, a
given theory can look
like a superstring theory in the perturbation theory for one coupling constant,
or like a
supergravity theory when expanded with respect to another. Perturbative
$p$-brane theories with $p>1$
do not seem to emerge. However, in each case the analysis depends on knowing
the relevant parts of the $p$-brane spectrum,
and in some cases our knowledge is incomplete and so the results presented are
provisional.
In particular, while our knowledge of BPS  $p$-brane states is fairly complete,
we know very little about non-BPS states, which may also exist in a
perturbative spectrum. In some cases, these are metastable in some
weak-coupling regime, but cannot be extrapolated to states for other values of
the coupling.
In most cases, a satisfactory picture emerges
in terms of BPS states alone, but in some cases it appears that it is the
metastable non-BPS states that are perturbative and govern the dynamics in some
coupling constant regime.

 The results suggest that
M-theory or $p$-brane theories such as the
$11$-dimensional supermembrane, if they exist, cannot be treated perturbatively
as $p$-brane
theories (indeed, the uncompactified $11$-dimensional
supermembrane has no coupling constant arising from a scalar expectation value)
but that when
expanded with respect to a coupling constant such as a compactification
modulus, the perturbative
theory that emerges is a superstring or supergravity theory.
Remarkably, in many supergravity
theories, we learn that the perturbative spectrum with respect to certain
coupling constants
includes strings, and so consistent perturbative quantization requires string
theory.
Conversely, expanding a superstring theory in a compactification modulus
typically leads to
a perturbative spectrum which is that of a supergravity theory
rather than a superstring theory.

These arguments identify some of the  states which should be used  in trying to
construct
a perturbation theory and which  should propagate in loops. Whether or not the
quantum perturbation theory \lq exists' is another matter.  For superstrings, a
consistent finite perturbation
theory is expected to exist, while for supergravity, it exists at best
in the presence of a cut-off.
At present, we do not have any good  definition of  what a perturbative
$p$-brane theory might be,  so it is just as well that no such theory appears
to arise in any strong coupling limit.
We also do not have any good definition of M-theory, but in the strong coupling
limits
that are naturally formulated in 11 dimensions, the perturbative states are
just the supergravity $0$-brane multiples and all other states of M-theory
appear non-perturbatively.
These results would be consistent with the existence of a non-perturbative
quantum M-theory which, when expanded perturbatively with respect to a
compactification modulus, gives
a perturbative superstring or superparticle theory.
Many of the results presented here can be viewed as giving further evidence
supporting the conjectures made in [\Witten] for the strong-coupling behaviour
of various string theories: the evidence given in [\Witten] was based on
studying particle states, and here it is shown that when $p$-brane states are
also considered, the conclusions remain the same.

\chapter{Low-Energy Effective Actions}

The low-energy effective action for the heterotic string in $D$ dimensions
includes the bosonic  terms
$$
\int
 d^{D} x \sqrt{-g} e^{-2 \Phi} \left( R+ 4(\partial   \Phi )^2 +{1 \over 2}
(\partial \chi)^2
- {1 \over 12}H^2
-{1 \over 4}F^2
\right)
\eqn\het$$
where $\Phi$ is the dilaton,  $\chi ^i$ are the remaining scalar fields with
sigma-model action
$ (\partial \chi)^2 = g_{ij}(\chi) \partial _ \mu \chi^i \partial ^\mu \chi^j$,
$H$ is the field strength for the two-form potential $b$ and $F$ is the
Yang-Mills field strength taking values in the
 gauge group $G= U(1)^{2(10-D)} \times K$ where $K$ is a group of  rank $26-D$.
 The same action
with the $F^2$ term omitted describes the NS-NS sector of the type II string in
$D$ dimensions. The
action \het\ describes the low-energy heterotic string for  small values of the
string coupling $g
\equiv \left\langle e^\Phi \right\rangle$ and we wish to find the dynamics for
strong coupling.  One
approach is to seek a change of variables to an action that might be
interpreted as the low-energy
effective action of a dual string or $p$-brane theory  with coupling constant
$g'= g^{- \beta}$ for
some constant $\beta$. If $\beta>0$, this would be consistent with the
conjecture that  the strong
coupling limit of the original theory is the weak coupling limit of the dual
theory, and {\it vice
versa}. As we shall see, this process can lead to more than one
  candidate for a dual theory, but it is nonetheless useful in locating
possible dual theories.

The rescaling
$$ g_{\mu \nu}\to e^{2 \alpha\Phi} g_{\mu \nu}, \qquad \Phi \to -\beta \Phi
\eqn\subs$$
with
$$ \alpha= -{2 \over D-6}, \qquad \beta = {4 \over D-6}
\eqn\alis$$
leads to the dual action
$$
\int
 d^{D} x \sqrt{-g} \left[ e^{-2 \Phi} \left( R+ {26-D \over 4}(\partial   \Phi
)^2 +{1 \over 2} (\partial
\chi)^2\right) - {1 \over 12}H^2  -{1 \over 4}e^{- \Phi} F^2 \right]
\eqn\open$$ For $D=10$ and $G=SO(32)$, \open\ is the low-energy effective
action for the type I
string, motivating the conjecture that the $SO(32)$ heterotic and type I
strings are dual
 [\Witten,\GrWitt].  For $ D>6$, $\beta >0$ and \het\ and \open\ are a possible
 dual pair of
theories, with the weak coupling regime of one corresponding to the strong
coupling regime of the
other. For
$D<6$, $ \beta<0$ and the weak coupling regime of  one corresponds to the weak
coupling of the
other. An equivalence between the corresponding two weakly coupled string
theories
   seems unlikely.
For $D=6$, the transformation \subs, \alis\ is singular.  The action \open\
would be the effective
action for a type I string in $D$ dimensions in which all of the group $G$
arose from Chan-Paton
factors, or from some generalisation in which the Yang-Mills kinetic term arose
at the same order
in perturbation theory as the disk diagram. This suggests the conjecture that
the heterotic string
in $D>6$  dimensions with gauge group $G$ is dual to  some generalised type I
theory with gauge
group $G$. In
$D=10$, this would require a new \lq generalised type I string' with $E_8
\times E_8 $ gauge group, and such a theory has recently been conjectured to
arise from
 an orbifold compactification of 11-dimensional M-theory [\Mwit]. For
$D<10$,
  this new type I theory would appear to be different from the theory obtained
by dimensionally reducing the
$D=10$  type I string, as this would lead to an action of the form \open\ with
gauge group
$G=SO(32)$ (possibly broken by Wilson lines) but with extra abelian gauge
fields from the closed
string sector: $10-D$ abelian graviphotons   with lagrangian
$e^{-2 \Phi } F^2$ and
$10-D$ abelian gauge fields  from the RR two-form with lagrangian  $  F^2$ that
do not couple to
the dilaton.

The rescaling \subs\ with
$$ \alpha= -{2 \over D-4}, \qquad \beta = {2 \over D-4}
\eqn\alisii$$
gives the action
$$
\int
 d^{D} x \sqrt{-g} \left[ e^{-2 \Phi} \left( R+  {10-D \over 4}(\partial   \Phi
)^2 +{1 \over 2} (\partial \chi)^2 \right)
- e^{2 \Phi}{1 \over 12}H^2
-{1 \over 4}F^2 \right]
\eqn\typii$$
and rewriting in terms of the dual $(D-4)$-form potential $\tilde b$ with
$(D-3)$-form field strength $\tilde H= d \tilde b$ given by the Hodge dual of
the field strength
$H$ of the two-form potential
$$e^{2 \Phi}*H = \tilde H
\eqn\hdual$$  gives
$$
\int
 d^{D} x \sqrt{-g} \left[ e^{-2 \Phi} \left( R+ {10-D \over 4}(\partial   \Phi
)^2 +{1 \over 2} (\partial
\chi)^2 -{1 \over 2 \,  (D-3)!}\tilde H^2
 \right) -{1 \over 4}F^2 \right]
\eqn\typiidu$$ As $\tilde H$ couples naturally to a $p$-brane with $p=D-5$, it
has been suggested
that this action be interpreted as the low-energy action for a $p$-brane theory
with $p=D-5$
[\DL,\StrFiv]. This suggests a duality between the heterotic  string and a
$p$-brane theory for
$D>4$ (so that the strong coupling limit of one is the weak coupling limit of
the other)
[\DL,\StrFiv]. For
$D=6$, this would be  a string/string duality, while for $D=10$ this would be a
string/five-brane
duality. Thus for $D=10$ we have (at least) two candidates for the dual of the
heterotic string: a
type I theory and a five-brane theory.

Finally, the
  rescaling \subs\ with
$$ \alpha= -{2 \over D-2}, \qquad \beta = {4 \over D-2}
\eqn\alisein$$ gives the Einstein frame action
$$
\int
 d^{D} x \sqrt{-g} \left[   R+ {2-D \over 4}(\partial   \Phi )^2 +{1 \over 2}
(\partial \chi)^2   - e^{-2
\Phi}{1 \over 12}H^2
-{1 \over 4}e^{- \Phi}F^2 \right]
\eqn\ein$$

Similar results apply for the low-energy effective actions of type II and type
I superstrings.
In particular, for type II strings the 2-form potential can be dualised to a
6-form potential
[\cham], so that a string theory effective action is replaced by a \lq 5-brane
effective action',
and various other anti-symmetric tensor gauge fields can be dualised [\cham].

\chapter{Coupling Constant Dependence in Four Dimensions}

In four dimensions, the toroidally compactified heterotic string at a generic
point in its moduli
space has gauge group $U(1)^{28}$ so that there are $28$ electric charges $q_I$
and $28$ magnetic
charges $p^I$ (as defined in [\HT]). The supergravity field equations are
invariant under
$K=SL(2;\R)\times O(6,22)$ and this is broken down to the integral subgroup
$SL(2;\Z)\times
O(6,22;\Z)$ of S and T dualities  by quantum effects [\SS,\Sen]. The scalar
fields take values in
the coset space
$${SL(2;\R) \over U(1)}\,  \times \,
{O(6,22)\over O(6)\times O(22)}$$
and the $56$ electric and magnetic charges transform as
the irreducible $({\bf 2}, {\bf 28})$ representation of
$SL(2;\R)\times O(6,22)$. The $N=4$ supersymmetry algebra has $6$ electric and
$6$ magnetic central
charges $\tilde q_I , \tilde p^I$ given in terms of the charges
$q_I,p^I$ by $\tilde q = K q, \tilde p = K p$ where $K$ is a $6\times 28$
matrix function  of the
moduli that are given by the asymptotic values of the scalar fields taking
values in
$O(6,22)/O(6)\times O(22)$.
The  $N=4$ Bogomolnyi mass formula for BPS saturated states (preserving half
the supersymmetry) is
[\SS,\Sen]
$$
M^2 =
\pmatrix {\tilde p & \tilde q}
{\cal S}
\pmatrix{\tilde p \cr \tilde q}
\eqn\rhet
$$
where $M$ is the ADM mass in the Einstein frame and
$$
{\cal S} = { 1 \over \lambda _2}\pmatrix {{ \vert  \lambda \vert}^2 & \lambda
_1 \cr
\lambda _1 & 1}
\eqn\sis
$$
is an $SL(2,\R)$ matrix depending on $\lambda = \langle a +ie^{-2\Phi }\rangle
=\lambda _1 + i
\lambda_2$, where $a $ is the axion and $\Phi $ is the  dilaton.
(In the notation of [\SS,\Sen], $K^tK={1 \over 16}(M+L)$.)
For vanishing axion expectation value $\lambda_1$, the mass is given by
$$M^2 = g^2 \tilde q^2 + {1 \over  g^2} \tilde p
^2
\eqn\mass$$
where $g =\langle  e^{\Phi }\rangle$ is the string coupling constant.
For the mass $M_s$ measured with respect to the stringy metric $\tgmn$, which
is given in terms of
the Einstein metric $\gmn$ by $\tilde g_{\mu
\nu} =e^{2\Phi}
\gmn$, the formula \mass\ becomes
$$M_s^2 =   \tilde q^2 + \left( {1 \over  g^2} \tilde p \right)
^2
\eqn\masss$$
This  form was to be expected, since the electric charges are carried by
perturbative string
states while the magnetic ones arise from solitons, and  the mass of a
magnetically charged state
has the standard $1/g^2$  coupling constant dependence of a soliton.

The $SL(2,\R)$ symmetry acts as
$$
{\cal S} \to  \Lambda ^t{\cal S} \Lambda , \qquad \pmatrix{\tilde p \cr \tilde
q} \to \Lambda^{-1}
\pmatrix{\tilde p \cr \tilde q}
\eqn\sltran$$
where $\Lambda $ is a $2\times 2$ matrix in  $SL(2,\R)$, and $(p ,q)$
transforms in the same way
as $(\tilde p,  \tilde q)$.   The Einstein metric
$\gmn$ is invariant, and the mass formula \rhet\ is   manifestly invariant.
The $SL(2,\Z)$ transformation given by \sltran\ with
$$\Lambda = \pmatrix { 0 &  1 \cr
-1 & 0}\eqn\ertg$$
interchanges electric and magnetic charges while $\lambda \to - 1/\lambda$. If
$\langle a \rangle=0$, then  the coupling constant is inverted, $g \to 1/g$,
and weak and strong
coupling regimes are interchanged.
Thus the theory is self-dual: the strongly coupled regime can be treated using
perturbation theory
in the small coupling constant $\hat g = 1/g$
and this gives a dual heterotic string theory.
In the weakly coupled theory, the electric charges $q$ were carried by
$g$-perturbative states
(i.e. ones that arise in the perturbation theory with respect to $g$) and
the magnetic ones $p$ by solitons, while   in the dual theory the electric
charges $q$ are carried
by solitons and the magnetic ones $p$ are carried by states that arise as $\hat
g$-perturbative
states.

For the dual theory, it is appropriate to use the dual
 stringy metric $\tgmn$, which is given in
terms of the Einstein metric $\gmn$ by $\tilde g_{\mu
\nu} =e^{-2\Phi}
\gmn$. The
mass $M_d$ measured with respect to this metric is then, from
 \mass,
$$M_d^2 =   \tilde p^2 +
 \left( {1 \over \hat g^2} \tilde q \right)
^2
\eqn\massd$$
This is consistent with the fact that $p$ is carried by $\hat g$-perturbative
states and $q$ by
solitons for $g >>1$ ($\hat g << 1$).

Consider now the type II theory toroidaly compactified to four dimensions.
 The gauge group is again $U(1)^{28}$, so that there are again $28+28$ electric
and magnetic
charges that form a $56$-vector ${\cal Z}$ which in the quantum theory must
take values in a
self-dual lattice. The low-energy effective action is that of $N=8$
supergravity [\CJ], which has
an
$E_{7(7)}$ symmetry of the equations of motion which is conjectured to lead to
a discrete
$E_7(\Z)$ U-duality symmetry of the string theory [\HT].
The charge vector
 ${\cal Z}$
transforms according to the irreducible {\bf 56} representation of $E_7$, which
has
the decomposition
$$
{\bf 56} \rightarrow ({\bf 2},{\bf 12}) + ({\bf 1}, {\bf 32})
\eqn\btwo
$$
under the subgroup $SL(2;\R)\times SO(6,6)$.
This is to be
compared with the heterotic string, for which
   the charge vector $(p,q)$ has the decomposition
$$
({\bf 2}, {\bf 28}) \rightarrow ({\bf 2},{\bf 12}) + 16\times ({\bf 2},{\bf
1})
\eqn\bone
$$
in terms of representations of $SL(2;\R)\times SO(6,6)$.
In both cases there is a common sector corresponding to the $({\bf 2},{\bf
12})$ representation of $SL(2;\R)\times SO(6,6)$, plus an additional
32-dimensional representation corresponding, for the heterotic string, to the
charges for the additional $U(1)^{16}$ gauge group and, for the type II
strings, to the charges for the  RR sector gauge fields.

The scalar fields take values in the coset space
$E_7/SU(8)$ and
can be represented by a $56 \times 56$ matrix $\V$ that transforms under $E_7$
from the right and
under local $SU(8)$ transformations from the left [\CJ]. The charge vector
$\cal Z$ enters the
Bogomolnyi mass formula through the $E_7$-invariant combination
$\bar {\cal Z}= \overline \V {\cal Z}$, where $\overline \V $ is the asymptotic
value of $\V$.
The Lie algebra of $E_7$ can be decomposed into that of $SL(2;\R)\times
SO(6,6)$ and its orthogonal
complement $X$, so that $\overline \V$ can be written as $\overline V= STR$
where
$S \in SL(2;\R)$, $T\in  SO(6,6)$ and $ R$ is the exponential of an element of
$X$.
Then the dressed charge vector $\tilde {\cal Z}=TR {\cal Z}$ decomposes into
$12$ doublets of $SL(2,\R)$, consisting of $12+12$ \lq dressed' electric and
magnetic charges
$(\tilde p^I,
\tilde q_I)$, together with $32$ singlets of $SL(2;\R)$, the \lq dressed' RR
charges $\tilde r_a$.
 The ADM mass formula for  Bogomolnyi states in the Einstein-frame is then
$$
M^2 =
\pmatrix {\tilde p & \tilde q}
{\cal S}
\pmatrix{\tilde p \cr \tilde q}+ \tilde r^2
\eqn\rtwo
$$
where
$
{\cal S}$ is given in terms of $\lambda = \langle a +ie^{-2\Phi }\rangle$ by
\sis.
The dependence on $\lambda$ can be understood from group theory, and in
particular the
fact that the RR charges $\tilde r$ occur without any  dependence on
$\lambda$ follows from the fact that they are $SL(2,\R)$ singlets.
 For vanishing   $\lambda_1$, the mass is given by
$$M^2 = g^2 \tilde q^2 + {1 \over  g^2} \tilde p
^2 + \tilde r^2
\eqn\massii$$
For the
  mass $M_s$ measured with respect to the stringy metric $e^{2\Phi}
\gmn$, this becomes
$$M_s^2 =   \tilde q^2 + \left({1 \over  g }\tilde r  \right)^2+ \left({1 \over
 g^2} \tilde p
\right) ^2
\eqn\masssii$$
while for mass $M_d$ corresponding to the dual
 stringy metric $ e^{-2\Phi}
\gmn$ the  Bogomolnyi mass formula
   is
$$M_d^2 =   \tilde p^2 +  \left({1 \over \hat g }\tilde r  \right)^2+
\left({1 \over \hat g^2} \tilde q
\right) ^2
\eqn\massdii$$
Thus, as in the $N=4$ case, NS-NS electric and magnetic charges are
interchanged under
strong/weak duality, but the states carrying RR charges are non-perturbative at
both weak and
strong coupling. Whereas magnetic charges are associated with effects   with
the usual
non-perturbative coupling dependence of
$e^{-1/g^2}$, the RR charges are associated with ones with the stringy
dependence
$e^{-1/g}$, similar to that found in matrix models [\Shenker].

\chapter{Bogomolnyi States and $p$-branes}

The particle states with a given charge that saturate a Bogomolnyi bound are
expected to be in the
theory for all values of the coupling constants, although they might be
represented as elementary
excitations for some values of the  coupling constants and as solitons for
others.
For example, in $N=4$ supersymmetric Yang-Mills theory with gauge group
spontaneously broken to
$U(1)$, the Montonen-Olive duality conjecture implies that the magnetic charges
are carried by BPS
monopoles at weak coupling, while the same states extrapolated to strong
coupling are elementary
particles -- W-bosons -- of the dual $N=4$   Yang-Mills theory describing the
strongly coupled
theory. There is as yet no clear picture of what happens for intermediate
values of the coupling.
 With enough supersymmetry (at least $N=4$ in four dimensions) the Bogomolnyi
states are specified
by their charges and their masses are given in terms of these charges and the
coupling constants
(corresponding to scalar field expectation values) by the Bogomolnyi formula,
which is determined
 completely by considerations of low-energy supersymmetry.

Similarly, there are $p$-brane states in the theory that saturate a Bogomolnyi
bound and are also
expected to be in  the
theory for all values of the coupling constants, and again these can be
elementary for some values
of the couplings and solitonic for others. Here, we shall mainly be concerned
with states that
break half the supersymmetry. The $p$-brane states of the superstring theory
are
sources for the long range fields and so lead to field configurations
which are the $p$-brane
solutions of the low-energy effective  supergravity theory.
For example, most  superstring theories   (with the possible exception
of type I)  have a 1-brane saturating a Bogomolnyi bound and which is the
fundamental string of the
weakly coupled theory but a solitonic string of the strongly coupled dual
theory.  Most string
theories   also have Bogomolnyi
$p$-branes that, at weak coupling, are either non-singular solitons  of the
weakly coupled theory or
are associated with D-branes, and these should again extrapolate to
$p$-brane states at other values of the couplings.  The mass per unit
$p$-volume is given
by the Bogomolnyi formula. This is again determined by low-energy
supersymmetry, but it will be
convenient here to  read off the coupling constant dependence of the
$p$-density from that of a
$p$-brane solution of the effective field theory and extrapolate the formula to
all values of the
couplings. This gives the same result as the Bogomolnyi formula, but can also
be applied to
non-Bogomolnyi
  solitons, although in the latter case the lack of supersymmetry can allow
quantum corrections to
the masses and charges.

Consider an action of the form
$$
\int
 d^{D} x \sqrt{-g} \left(e^{-2 \Phi}  \left[ R+ 4(\partial   \Phi )^2 \right]
-e^{-  \gamma \Phi}{1 \over 2 (n+1)!}G^2 + \dots \right)
\eqn\gen$$
where $G$ is the $(n+1)$-form field strength for an $n$-form gauge potential
$C$ (the case $n=0$ is
a scalar field) and $\gamma$ is a constant.  Such actions arise as part of
superstring effective
actions. The constant $\gamma$ takes the values   $\gamma=2$ for the heterotic
string or for a NS-NS
field in a type II string, while $\gamma=0$ for RR fields of type I and II
strings and $\gamma=1$
for
 the type I vector fields (and type I  scalars in $D<10$) whose kinetic terms
arise from a disk diagram.
Such theories have $p$-brane solitons which couple to $C$ and   carry
corresponding charges [\HS,\DL].
These consist of electrically charged $(n-1)$-brane solitons
and magnetically charged $(D-n-3)$-brane solitons.
It is straightforward to check (cf [\DL])  that the mass per unit $p$-volume
$M_p$ of a $p$-brane
soliton of this type  has the following coupling constant dependence:
$$
M_p \sim  g^{- \epsilon_p}T_p, \qquad \epsilon_p = \cases{
  1-{\gamma \over 2}& for electric $p$-branes \cr
   1+   {\gamma \over 2}& for magnetic $p$-branes }
\eqn\masbran$$
Here
  the coupling constant $g$ is  given by the asymptotic value of $e^\Phi$
(which is assumed constant) and $T_p$  is a dimensionful constant that can be
thought of as an
effective brane tension, and which contains the dependence on any coupling
constants other than $g$.
In the case $n=0$, $C$ is a scalar field which  formally appears to couple to
an electric $p$-brane with  $p=-1$. This can be interpreted as an
instanton in the Euclidean version of the theory [\Dbranes,\sevbrane], but this
case will not be
considered further here as we will be interested in the $p$-brane spectrum for
Lorentzian signature.
The scalar also couples to a magnetic $(D-3)$-brane (e.g. a string in $D=4$ or
a 7-brane in $D=10$
[\sevbrane]) and these will be considered here.

Configurations satisfying $p$-brane boundary conditions typically satisfy a
Bogomolnyi bound of the
form $M_p \ge \vert {\cal Z}  \vert $ where  $M_p$ is the mass per unit
$p$-volume and $ \vert
{\cal Z}
\vert $ is the electric or magnetic central charge, which is given in terms of
the asymptotic
values of the scalar fields and the magnetic charge  $P=\int _{S^{n+1}}
 G$ (given by an integral over an $(n+1)$-sphere  surrounding the
$(n-1)$-brane) or the electric
charge $Q=\int _{S^{D-n-1}}
 e^{-  \gamma \Phi}*G$ (integrated over a $(D-n-1)$-sphere surrounding the
$(D-n-3)$-brane).  The
precise coupling constant dependence of the relation between ${\cal Z}$ and
$Q,P$ is determined by
supersymmetry, and the dependence on scalars other than $\Phi$ has been
suppressed.
 Of particular interest are those
$p$-brane solitons that break half the supersymmetry, and consequently saturate
the Bogomolnyi
bound  so that the density is given by
$M_p= \vert {\cal Z} \vert $ [\DGHR,\DL]. This takes the form \masbran, but in
this case the
relation is
 expected to be exact so
that the density of the corresponding quantum state is given by \masbran.

The mass formula \masbran\ is valid for the string metric $\gmn$,
but it is straightforward to generalise to an arbitrary metric $\tgmn$ given by
$$
\tgmn= e^{2 \alpha \Phi} \gmn
\eqn\resc$$
Lengths and masses as measured in the asymptotic region (where $e ^\Phi \sim
g$)
with respect to the two metrics
are related by
$\tilde L = g^ \alpha L$, $ \tilde M = g^{- \alpha}M$ so that the masses per
unit
$p$-volume  are related by
$$ \tilde M_p = g^{-(p+1) \alpha} M_p
%\sim g^{- [\epsilon_p+ (p+1) \alpha]}T_p
\eqn\mext$$
 Then for a general metric $\tgmn$ \masbran\
becomes
$$
M_p \sim  g^{- [\epsilon_p+ (p+1) \alpha]}T_p, \qquad \epsilon_p = \cases{
  1-{\gamma \over 2}& for electric $p$-branes \cr
   1+   {\gamma \over 2}& for magnetic $p$-branes }
\eqn\bmas$$
For $\alpha= -2/(D-2)$, $\tgmn$ is the Einstein metric and \bmas\ reproduces
the results of [\DL].

The $g$-dependence of the masses clearly depends on the choice of metric.  In
the string metric
$\alpha =0$, the ratio $M_p/T_p$ is independent of $g$ for the fundamental
string of some string
theory ($\epsilon_p=0$ and $\gamma =2$), while for the corresponding magnetic
$(D-5)$-brane
soliton the ratio $M_p/T_p$ is proportional to $g^{-2}$ and for RR branes it is
proportional to
$g^{-1}$. Thus in the limit $g \to 0$, the solitons and D-branes become
infinitely massive and
decouple, leaving  the fundamental string in the perturbative spectrum. This is
precisely what is
required for a sensible perturbation theory in some small coupling constant
$\lambda$ (which might be the string coupling or be associated with the
expectation of some
other scalar): some states with masses that are independent of
$
\lambda$ and which are the perturbative states of the theory, with all other
states having masses
that tend to infinity as
$\lambda \to 0$ so that they are non-perturbative states. This can now be
applied to the  strong
coupling limit. As we shall see, for any supergravity or superstring theory
there is a unique choice
of rescaled metric \resc\ such that the ratio
$M_p/T_p$ in that metric is independent of $g$ for some states and tends to
zero as $g \to \infty$
for all other states, so that they decouple in the strong coupling limit. There
is then a sensible
perturbation theory in $\hat g = 1/g$ with the perturbative states consisting
of those with
$g$-independent masses in the \lq dual metric'.  Note that the set of $\hat
g$-perturbative
states  identified in this way is {\it not} the same as the set of
non-perturbative states of the
original string theory, i.e. those states for which the string-metric ratio
$M_p/T_p$ tends to infinity as
$\hat g
\to 0$. More generally, as we shall see, given any supergravity or superstring
theory and any
choice of coupling constant $\lambda$, there are precisely two preferred
rescaled metrics, one
which leads to a sensible non-trivial weak-coupling perturbation theory in $
\lambda$ of the type described above, and one which leads to a sensible
non-trivial strong-coupling
perturbation theory in
$ 1/\lambda$.
Of course the theory can be studied in whichever metric one chooses. The point
here is that   the
preferred metrics are the ones in which the perturbation theory looks most
natural, with
the perturbative states having $\lambda$-independent masses.
 It is conceivable that other choices of metric might lead to consistent
perturbation theories in this way,
 but these would be non-maximal, as the perturbative spectrum would
be strictly smaller than that arising from one of the two preferred metrics, as
will be
seen in the examples below.

In this way, it is possible to identify the part of the perturbative spectrum
consisting of
BPS-saturated $p$-branes that emerges at strong coupling and at weak coupling
in $lambda$ for any
given theory.  We shall first check that this gives the expected results in
cases where U-duality
gives an alternative derivation of strong-coupling dynamics. It will    then be
applied   to other
cases, and    in particular to those theories described in the introduction,
where it will help
decide between different conjectures for the strong-coupling dynamics.

\section {Type II in Four Dimensions}

Consider the example of the four-dimensional type II string (compactified on a
torus) for which the
masses of Bogomolnyi
$0$-branes is given by
\masssii\ in the stringy metric.
The $g$-perturbative states are those carrying NS-NS electric charge $\tilde q$
and have
$g$-independent masses $M_s = \tilde q$, while the non-perturbative states
carrying NS-NS
 magnetic or RR charges
have masses
$M_s = g^{-2} \tilde p$ and $M_s = g^{-1} \tilde r$
respectively and become massless at strong coupling, $M_s \to 0$ as $g \to
\infty$.
 However, the
NS-NS magnetic states become massless fastest and  it is these that
become the $\hat g$-perturbative
states at strong coupling ($\hat g =1/g$). For a general metric \resc,
\masssii\ becomes
$$M ^2 =  \left[ g^{-  \alpha}\tilde q \right]^2 + \left[ g^{-(1+\alpha)}\tilde
r \right] ^2 +
\left[ g^{-(2+\alpha)}
\tilde p  \right] ^2
\eqn\masssiialp$$
so that for states carrying only one type of charge, the masses are
$M=g^{- \alpha}\tilde q$, $M=g^{-(1+\alpha)}\tilde r$ or $M=g^{-(2+\alpha)}
\tilde p$.

There are two preferred values of $\alpha $ which give good perturbation
theories.
The stringy metric $\alpha =0$ gives the perturbative spectrum described above.
The
  value $\alpha =-2$
gives the \lq dual stringy metric' for which NS-NS magnetically charged states
have $M= \tilde p$ and are perturbative, while  electric states have
$M=\hat g^{- 2}\tilde q$ and RR ones have $M=\hat g^{-1}\tilde r$.
States carrying $\tilde q$ or $\tilde r$ charge (including those with more than
one type of charge)
have masses that
diverge in the strong-coupling limit $\hat g =1/g \to 0$ and so are
non-perturbative in
the strong-coupling perturbation theory.
For metrics with
$-2 < \alpha <0$
there are   states that become massless as either $g \to 0$ or $g \to \infty$
so that there is no
consistent weak or strong coupling perturbation theory.
In particular,  for
the Einstein metric $\alpha =-1$   the RR states have $g$-independent masses
but  there is no
consistent weak or strong coupling perturbation theory
 based on the RR states and neutral states alone.
For   $\alpha>0$, the Bogomolnyi formula \masssiialp\ implies that
all charged states have masses that depend on a negative power of $g$,
so that they decouple in the $g \to 0$ limit and any states with
$g$-independent masses must be
uncharged. Thus a weak-coupling perturbation theory, if consistent, would be
non-maximal as it would
contain no perturbative charged states,
and would correspond to the limit of the usual perturbation theory in which the
unit of charge is
taken to infinity.   In the strong coupling limit with $\alpha>0$, all charged
states become
massless so that it would not make sense to base a perturbation theory on
uncharged states alone.
  Similar considerations apply for $\alpha <-2$, for which there is a strong
coupling perturbation
theory involving no charged states. This leads us to the two preferred metrics
with
$\alpha =0,-2$
and the corresponding weak and strong coupling perturbation theories.
This is precisely the result predicted by U-duality: the theory is self-dual,
with the strong
coupling  theory given by the a dual version of the original theory with NS-NS
electric and magnetic charges interchanged.

\section{Type IIB in $D=10$}

Type IIB supergravity and the type IIB superstring in $D=10$ both have two
$1$-brane solutions,
two $5$-brane solutions and a self-dual $3$-brane solution [\HS], together with
a 7-brane and a
9-brane [\Dbranes,\sevbrane]. (We will only be interested in the Lorentzian
signature solutions
here and so will not consider the instantonic $-1$ brane [\sevbrane].) The
values of
$p,
\gamma,
\epsilon _p
$ and
$\rho_p
\equiv
\epsilon _p /p+1$ are given in the following table:

\vskip 1cm
\vbox{
\begintable
 Value of $p$ | $1 $(NS-NS) |$ 5$ (NS-NS) | $1'$ (RR) | $5'$  (RR) | $3^+$
(RR) \elt
 $\gamma$ | 2 | 2 | 0 | 0 | 0 \elt
 $\epsilon_p$ | 0 | 2 | 1 | 1 | 1 \elt
 $\rho_p= \epsilon_p/p+1$ | 0 | 1/3 | 1/2 | 1/6 | 1/4 \elt
 String metric $M_p/T_p$  | 1 | $g^{-2}$ | $g^{-1}$ | $g^{-1}$ | $g^{-1}$ \elt
 Dual string metric $M_p/T_p$   | $g$ | $g $ | 1 | $g^{2}$ | $g $ \elt
 Einstein metric $M_p/T_p$  | $g^{1/2}$ | $g^{-1/2}$ | $g^{-1/2}$ | $g^{1/2}$ |
1 \elt
 General metric $M_p/T_p$    | $g^{-2\alpha}$ | $g^{-6(1/3+\alpha)}$ |
$g^{-2(1/2+\alpha)}$ |
$g^{-6(1/6+\alpha)}$ | $g^{-4(1/4+\alpha)}$
\endtable

\centerline{{\bf Table 1} Type IIB Brane-Scan.}}

%\catcode`\|=12
 \vskip .5cm
The 1-branes are electric, the 5-branes are magnetic and the self-dual 3-brane
has equal electric and
magnetic charges. Those with $\gamma=2$ couple to the NS-NS 2-form potential
while those with
$\gamma =0$ couple to RR gauge fields.
(There is no action of the form \gen\ for the 4-form potential with self-dual
field strength, but
the mass formula \masbran\ can nonetheless be applied to the 3-brane with
$\gamma=0$ [\DL].)
This can be extended to include the 7 and 9 branes. It follows from [\Dbranes]
that the RR $p$-branes
with
$p=1,3,5,7,9$ all have
  $M_p/T_p$ given by $g^{-1}$ in the string metric and so by $g^{-
[1+(p+1)\alpha]}$ with respect to the general metric, corresponding to
$\rho_p=1/p+1$.

In each case, the $p$-brane  mass
satisfies
$$M_p \sim g^{-(p+1)[\rho _p + \alpha]}T_p
\eqn\missb$$
and for a given $p$-brane, this is $g$-independent for the metric \resc\ with
$\alpha =- \rho _p$.
For each such $p$-brane metric except those the preferred cases $\alpha = 0,
-1/2$, some of the
soliton masses grow with
$g$ and others grow with $1/g$, so that there is no sensible weak or strong
coupling perturbation
theory;  this is seen explicitly in the table for the Einstein metric ($\alpha
=-1/4$) for
which the self-dual 3-brane has $g$-independent mass.
The special cases $\alpha = 0, -1/2$  are
 the minimum and maximum values of $\rho _p$ and
give masses that depend   on $g$ raised to a non-positive power or to a
non-negative one,
respectively. For
$\alpha =0$,
  the $g$-perturbative Bogomolnyi spectrum consists of the  1-brane with
$\gamma =2$, which is the
fundamental string so that we learn that the perturbation theory is that of a
string theory.
Note that even
if we had started with the IIB {\it supergravity} theory (including $p$-brane
solitons), we would
have learned that the weakly coupled theory is a perturbative {\it superstring}
theory, in which the
$\gamma =2$ soliton is identified with the fundamental string. All other
$p$-brane densities depend
on
$g$ to a negative power and so correspond to non-perturbative states whose
(string metric) densities
tend to infinity at weak coupling.

For strong coupling, we learn that the appropriate metric for describing the
dual theory is \resc\
with
$\alpha =-1/2$, and that the $1'$-brane which couples to the $\gamma =0$ 2-form
becomes  $\hat
g$-perturbative and is the fundamental string for the dual theory, while the
other $p$-branes become
$\hat g$-non-perturbative states of the dual theory. Thus the dual theory is
again a string
theory. This is in complete agreement with the conjecture that the type IIB
string has an $SL(2;\Z)$
duality symmetry [\HT] and so is self-dual: the strong coupling limit is a dual
type IIB
string theory in which NS-NS and RR charges have been interchanged. Thus
although in the original
string-metric, the NS-NS 5-brane has a mass which tends to zero as $g^{-2}$ at
strong coupling and
the  RR branes have masses that tend to zero as $g^{-1}$, it is the RR string
that here wins out over
the others and dominates the strong coupling theory, so that the theory is
self-dual rather than
dual to e.g. a 5-brane theory.

Mass formulae  similar to those given in the last section for the $D=4$
heterotic
string, which is also self-dual with an $SL(2;\Z) $ symmetry, can be written
down for the type IIB
theory. A 1-brane can carry an  NS-NS   charge $q$ (coupling to the NS-NS
2-form) or
a RR   charge $r$ (coupling to the RR 2-form)
or both, in which case it is \lq dyonic'. (Dyonic solutions can be constructed
from the NS-NS
string and RR string given in [\Oopen] by acting with $SL(2;\Z)$.) The Einstein
frame mass per unit
length
$M_1$ satisfies a Bogomolnyi bound which, for vanishing pseudo-scalar, is
$$
M_1^2 \ge g^{ } q^2 + g^{- 1} r^2
\eqn\sgffd$$
Strings saturating this bound have masses saturating this bound have masses
satisfying
$$
M_1^2=   q^2 + g^{-2} r^2
\eqn\ghjl$$
in the string frame
or
$$
M_1^2=   r^2 + {\hat g}^{-2} q^2
\eqn\ghgfl$$
in the dual string frame.

\section{The General Case}

Consider now an arbitrary supersymmetric theory in $D$ dimensions, which might
be a supergravity,
superstring or super-$p$-brane theory. Usually, we will restrict ourselves   to
situations in
which there is enough supersymmetry for the classical Bogomolnyi formulae to be
reliable in the
quantum theory. Let $\phi$ be some scalar field, which might be the dilaton
$\Phi$, or $-\Phi$, or
some compactification modulus. Then $g=e^{\langle\phi  \rangle}$ is one of the
coupling constants
of the theory and one can attempt to define perturbation theory with respect to
$g$.
 There is a spectrum of $p$-branes, and for each $p$-brane there is a constant
$\tilde
\rho _p$ such that the mass
per unit $p$-volume  has a dependence on the coupling constant $g$ that is
given with respect to
the Einstein metric by
$$M_p \sim g^{-(p+1)\tilde \rho _p }T_p
\eqn\misso$$
The dependence on all other coupling constants is absorbed into the effective
tension $T_p$.
In  string theory, the usual weak coupling perturbation expansion is that in
$\langle e^\Phi
\rangle$, but expanding in other couplings such as $\langle e^{-\Phi }\rangle$
or one of the
compactification moduli can give new insights and can lead to a new
perturbative theory,
which might be another string theory  or a
supergravity theory  or perhaps something more exotic such as a $p$-brane
theory.
(The 11-dimensional supergravity  or supermembrane theory has no scalars or
coupling constants and
so this analysis does not apply. If the 11-dimensional theory is compactified,
however, one of
the compactification moduli can be used as a coupling constant and a mass
formula \misso\ can be
found.)

If the Einstein metric is conformally rescaled to
$$g'{}_{\mu \nu}  = e^{2\tilde \alpha \phi} \gmn
\eqn\gresc$$
 then since the conformal factor
  tends to $g^{2\tilde\alpha  }$ in
the asymptotic region,   the mass formula becomes
$$M_p \sim g^{-(p+1)[\tilde \rho _p +\tilde \alpha]}T_p
\eqn\missr$$
with respect to the rescaled metric.
The weak coupling regime $g << 1$ is dominated by the brane or branes for which
$\tilde\rho _p$
takes its minimum value $\tilde\rho _{min}$ and the appropriate metric is that
given by \gresc\
with
$\tilde\alpha =-\tilde\rho _{min}$. The branes with $\tilde\rho _p=\tilde\rho
_{min}$ are then the
$g$-perturbative states, while the others are $g$-non-perturbative. For
example, if the only brane
with
$\tilde\rho _p=\tilde\rho _{min}$ is a 1-brane, then the weakly coupled theory
is a string theory,
while if the only such states are 0-branes, then the perturbative theory is a
supergravity theory.
Similarly, the strong coupling regime
$\hat g << 1$ with
$\hat g=1/g$ is dominated by the brane or branes for which
$\tilde\rho _p$ takes its maximum value $\tilde\rho _{max}$ and the appropriate
metric is that given
by
$\tilde\alpha =-\tilde\rho _{max}$, and again this might be a string theory or
a $p$-brane theory or
a field theory or a theory of a set of coupled $p$-branes with various values
of $p $.

It will be useful to define
$$\rho_p= \tilde \rho_p + {2 \over D-2}, \qquad \alpha= \tilde \alpha - {2
\over D-2}
\eqn\rtiis$$
so that \misso\ becomes
$$M_p \sim g^{-(p+1)[\rho _p + \alpha]}T_p
\eqn\miss$$
Whereas $\tilde \alpha =0$ corresponds to the Einstein metric, $\alpha =0$
corresponds to the
\lq string metric' for the \lq dilaton'   $\phi$.
Clearly  the maximum and minimum values of $\rho_p$ correspond to
the maximum and minimum values of $\tilde \rho_p$.

 In the
remainder of this paper, this structure will be explored in various theories.
In each case, there are
two preferred metrics    corresponding to
$\tilde \alpha = - \tilde\rho _{min}$ and $\tilde\alpha = -\tilde \rho _{max}$
(or equivalently
$\alpha = - \rho _{min}$ and $\alpha = - \rho _{max}$)   for weak and strong
coupling respectively and the corresponding perturbative spectrum can be read
off in each case.
 This will enable us to decide between various alternative
duality conjectures, such as those discussed in the introduction. Before
proceeding, however, a
number of remarks  are in order.

Firstly, the $p$-branes fit into supermultiplets and the formula \miss\ applies
to the whole
supermultiplet, although we shall usually refer explicitly to only the  member
of the
supermultiplet with lowest spin.
Secondly, given a $p$-brane with $M_p\sim g^a T_p$ for some $a$, there will
be further states
with
$M_p\sim n g^a T_p$ for   integers $n$ and these are degenerate in mass with a
configuration of $n${}  $p$-branes which each have $M_p\sim g^a T_p$.
Often in what follows, we will take
$T_p$ to be the minimum value, corresponding to elementary solitons, and not
consider the
\lq bound states' with   $M_p=nT_p$ explicitly.
Thirdly, the argument given above will identify only the charged massive BPS
states in the
perturbative spectrum. There will also be neutral states in the perturbative
spectrum, and in
particular these can include   particles  with zero mass and $p$-branes with
zero density.
In some cases, there are non-BPS states that are metastable   perturbative
states in some coupling constant regime, even though they cannot be continued
to states at other values of the coupling.
When these are included in the $p$-brane spectrum, they can
\lq win' over other $p$-branes and so need to be taken into account. However,
in most cases, they lose out to   BPS states and in those cases  will not be
considered explicitly.

\chapter {Ten and Eleven Dimensional Examples}

\section{Type IIA in $D=10$}

As discussed in the introduction, there are at least two rival proposals for
the strong coupling
limit of the
type IIA string, a 5-brane theory or an 11-dimensional theory, and the analysis
here will enable
us to decide between them.
 Type IIA supergravity and the type IIA superstring both have a 1-brane and a
5-brane
(coupling to the 2-form in the NS-NS sector of the string) plus $p$-branes with
$p=0,3,4,6,8$
(coupling to RR fields
of the string). The relevant properties are summarised in the following table.

\vskip 1cm
\vbox{
\begintable
 Value of $p$ | 1 | 5 | 0 | 2 | 4 | 6 \elt
 $\gamma$ | 2 | 2 | 0 | 0 | 0 | 0 \elt
 $\epsilon_p$ | 0 | 2 | 1 | 1 | 1 | 1 \elt
 $\rho_p= \epsilon_p/p+1$ | 0 | 1/3 | 1 | 1/4 | 1/5 | 1/7 \elt
 $M_p/T_p$ in string metric | 1 | $g^{-2}$ | $g^{-1}$ | $g^{-1}$ | $g^{-1}$ |
$g^{-1}$ \elt
 $M_p/T_p$ in dual string metric | $g$ | $g $ | 1 | $g^{3}$ | $g^{4}$ | $g^{6}$
\elt
 $M_p/T_p$ in Einstein metric | $g $ | $g $ | $g^{-1/2} $ | $g $ | $g^{3/2}$ |
$g^{5/2}$ \elt
 $M_p/T_p$ in general metric | $g^{-2\alpha}$ | $g^{-(2+6\alpha)}$ | $g^{- (1
+\alpha)}$ | $g^{-(1+4\alpha)}$ | $g^{-(1+5\alpha)}$ | $g^{-(1+7\alpha)}$
\endtable

\centerline{{\bf Table 2} Type IIA Brane-Scan.}}
%\centerline{}

%\catcode`\|=12

 \vskip .5cm
\noindent
In addition, there is an 8-brane solution of the massive type IIA theory
[\bergtown], which
formally has $M_p/T_p=g^{-1}$ in  the string metric [\Dbranes], corresponding
to
$\rho_8=1/9$. The minimum value of $\rho$ is $\rho _{min}=0$, corresponding to
the 1-brane which
becomes the fundamental string of the weakly coupled perturbative theory, for
which the appropriate
metric is the stringy metric with
$\alpha =0$. This is just as expected. The other $p$-branes all have densities
that tend to zero
as $g
\to
\infty$, so that arguments similar to those of [\Witten] might suggest that all
could be important for the strong coupling dynamics.
However, as $\rho _{max}=1$, the $0$-brane multiplet dominates at strong
coupling,
which is then a $\hat g$-perturbative supersymmetric field theory.
The BPS-saturated $p$-branes occuring in the strong coupling perturbative
spectrum
are then the
the 0-branes, which fit into short
massive   IIA supermultiplets with spins ranging from
 0 to  2. There is such a multiplet with
  electric charge $nq_0$
and mass $M= \vert nq_0\vert /g^{1+\alpha}$ for each integer $n \ne 0$, where
$q_0$
is some unit of charge  [\Witten]. In addition, there is the neutral massless
type IIA supergravity
multiplet (corresponding to $n=0$) and these  constitute the fundamental
excitations of the dual
theory. Thus the
states that emerge in strong coupling perturbation theory are those of the
$D=10$ type IIA
supergravity theory, plus an infinite tower of short  massive multiplets. As we
shall see in the next
section, this is  the perturbative spectrum that emerges from  11-dimensional
supergravity
compactified on a circle, when the radius of the compact dimension is treated
as a coupling constant.

\section{11-Dimensional Supergravity, M-Theory and  Supermembranes}

11-dimensional supergravity has a 2-brane [\DS] and a 5-brane soliton [\Gu],
but there are no
scalar fields whose expectation values can be used as coupling constants. Thus
there is no
perturbation theory in 11 dimensions that can be used to apply the analysis
developed above.
However, if the theory is compactified, then the moduli of the compactification
space can be used
as coupling constants. For example consider compactification on a circle. The
radius $R$ can be
used as a coupling constant: the massless spectrum consists of a D=10 type IIA
supergravity
multiplet, and there are in addition Kaluza-Klein momentum modes with mass $M
\sim q_0n/R$ for
integers $n$ and some constant $q_0$ (with respect to the Kaluza-Klein metric
[\Witten]).  These fit
into massive supermultiplets that saturate a Bogomolnyi bound and have spins
ranging from 0 to 2, and
have masses that tend to zero in the large
$R$ limit.  These have magnetic   partners, which arise as generalised
Kaluza-Klein monopoles in 11
dimensions and lead to 6-brane solutions on reducing to the 10-dimensional
theory [\PKT]. In
addition, the 11-dimensional 2 and 5 branes give rise to 10-dimensional 1,2,4
and 5 branes. This
gives rise to precisely the   type IIA brane scan given in table 2 after a
general
rescaling of the metric, with
$g = R^{3/2}$, with the Kaluza-Klein metric corresponding to $\alpha =-1/3$.
Thus for large $R$,
the coupling constant
$g = R^{3/2}$ is large and the 0-brane multiplet dominates. Then    the
perturbative states consist
of a massless type IIA supergravity multiplet, plus    an infinite Kaluza-Klein
tower of charged
ones.

Consider now the behaviour for small $R$.
For the compactified 11-dimensional {\it supergravity} alone without
$p$-branes, the
Kaluza-Klein states with $M\sim q_0n/R$ are non-perturbative in the expansion
in $R$, and the
resulting perturbative theory is of course the type IIA supergravity theory in
10 dimensions.
However, if the
$p$-branes are included in the spectrum, we see that
 at small $R$ the 1-brane
dominates and the perturbation theory is that of the type IIA string theory, so
that the
11-dimensional theory compactified on a small circle should be treated as a
string theory, not a
field theory.
If there is an 11-dimensional supermembrane or M theory  whose low energy
effective action is
 11-dimensional supergravity, it should also have a 2-brane and
a 5-brane soliton. On compactifying to 10 dimensions, the brane-scan would then
again be that of
table 2, so that the small $R$ limit would give the perturbative states of
type IIA string theory,
while the large $R$ limit would give type IIA supergravity.

If instead the theory is compactified on an $n$-torus of volume $V$, the
perturbative BPS states that
emerge in perturbation theory in $1/V$  consist of $n$ $0$-brane multiplets in
$11-n$
dimensions, which transform as an {\bf n} under the action of $SL(n)$. Each of
these multiplets
is the base of a Kaluza-Klein tower. The mapping class group of the torus is
the discrete subgroup
$SL(n;
\Z)$ and this is part of the U-duality group of the compactified theory.

\section{The Type I String and Heterotic Strings in $D=10$}

Consider the ten-dimensional $N=1$ supergravity theory coupled to $SO(32)$
super-Yang-Mills, which
is the low-energy effective field theory of both the type I string and the
$Spin(32)/\Z _2$ heterotic string theories.
The heterotic string form of the action \het\ is related to the type I form
\open\  by the field
redefinition \subs,\alis. This includes a dilaton sign flip, so that the strong
coupling limit of one effective field theory is the weak coupling limit of the
other [\Witten].
The theory has an electric 1-brane and magnetic 5-brane solution which couple
to $\bmn$ and
saturate a Bogomolnyi bound. There are also electric 0-branes and magnetic
6-branes coupling to
Yang-Mills fields taking values in a Cartan sub-algebra of $G=SO(32)$ [\HS],
but these do not
preserve any supersymmetries so that their status is unclear.
The 1-brane is the fundamental heterotic string and has a mass per unit length
which is independent
of
$g=e ^{\langle \Phi \rangle}$ in the natural heterotic string metric. The
0-branes also have
$g$-independent classical masses with respect to the heterotic string metric,
although these may
acquire quantum corrections.
The perturbative spectrum of the heterotic string then consists of the string
and possibly the
0-branes, corresponding to the minimum value of $\rho$.
The maximum value of $\rho$ for these $p$-branes occurs for the 5-brane, so
that if there are no
 $p$-branes in the theory other than those listed above, this would suggest
that the strong
coupling limit of the heterotic string is a perturbative 5-brane theory.
However, there is reason
to believe that there may be other $p$-branes in the theory that have higher
values of $\rho _p$,
and if this were the case these would be the strong coupling perturbative
states instead of the
5-branes.

The perturbative type I string has open and closed fundamental  string states.
These strings can break and so  are not stable and should not be expected to
saturate any BPS bound. However, these give metastable states that should be
included in the spectrum at weak coupling. A macroscopic type I string would
break, although
an unstable  solution  of the supergravity theory representing the field
configuration outside   a
macroscopic type I string could still exist, at least in the weak coupling
limit.
However,
as the type I string is not BPS saturated, there is no reason to expect that
such states could be extrapolated to strong coupling. Indeed, type I strings
become more likely to break as the coupling increases, so that such states
should not   exist at strong coupling, and there should be no corresponding
states in the weakly coupled   heterotic string.
Presumably, such a  fundamental   type I
string   should have mass/length that is independent of $g$ in
the type I string metric.
 The mass/length  of this type I string would then be
$M_1 \sim  g^{-1}T_1 $  in  the heterotic string metric and $M_1 \sim
g^{-1-2\alpha}T_1 $ in a
general metric given by scaling the heterotic metric by $e^{2\alpha
\Phi}$.
  The brane-scan for
all these $p$-branes is

\vskip 1cm
\vbox{
\begintable
 Value of $p$ | 1 (type I) |   1  | 5  | 0|6 \elt
 $\epsilon_p$ | 1 | 0 | 2 | 0|2 \elt
 $\rho_p= \epsilon_p/p+1$ | 1/2 | 0 | 1/3 | 0| 2/7 \elt
 $M_p/T_p$ in heterotic string metric | $g^{-1}$  | 1 | $g^{-2}$ | 1|$g^{-2} $
\elt
$M_p/T_p$ in type I string metric | 1 | $g^{1}$ | $g^{1}$ | $g^{1/2}$ |
$g^{3/2}$ \elt
$M_p/T_p$ in 5-brane metric | $g^{-1/3}$ | $g^{2/3}$ | 1 | $g^{1/3}$ |
$g^{1/3}$ \elt
$M_p/T_p$ in general metric | $g^{-(1+2\alpha)}$ | $g^{-2\alpha}$ |
$g^{-(2+6\alpha)}$ |
$g^{-\alpha}$ |
$g^{-(2+7\alpha)}$
 \endtable

\centerline{{\bf Table 3} Heterotic/Type I  Brane-Scan.}}

 \vskip .5cm
\noindent This table also includes the coupling constant dependence in the
so-called 5-brane metric
given in terms of the heterotic string metric  by \resc\ with $\alpha =-1/3$.
Note the unusual
$g$-dependence of the $0$ and $6$ branes in the type I metric.

The maximum and minimum values of $\rho$ are indeed those corresponding to the
type I and
heterotic strings and the perturbative heterotic and type I strings are a dual
pair with
the strong coupling regime of one corresponding to the weak coupling regime of
the other.
 There may be other $p$-branes of the theory, but it seems unlikely that they
will have
$\rho<0$ or $\rho>1/2$.
 Indeed, if there were $p$-branes with such values of $\rho$, then it seems
likely that there would
be problems for
 the perturbative formulation of either the heterotic or type I string, and no
such difficulties
are apparent.

If   the metastable type I string were not   included in the brane-scan   in
this way for weak type-I coupling, then it would be hard to find a plausible
duality picture unless there were some other as yet unsuspected $p$-brane
states of the theory. For example, if there was no type I string states  in the
heterotic string at strong coupling and
no other states with $\rho <1/6$, then table 3 would support the  conjecture
that the
strong coupling limit of the heterotic string is a perturbative  5-brane
theory, and presumably
the weakly coupled string would correspond to the strongly coupled 5-brane
[\DL,\StrFiv].  However,
table 3 would still support the conjecture that the strong coupling limit of
the type I string is a
weakly coupled heterotic string, and this would then imply the implausible
statement that the weakly coupled type I string is equivalent to   weakly
coupled perturbative
5-brane theory.
One way out of this would be if the heterotic string solution of the type I
theory [\Oopen,\dab]
were not to be regarded as an acceptable soliton solution (perhaps because of
its singularity
structure [\Polwit]) but it has recently been shown that the heterotic string
indeed arises as a
D-brane in the type I string theory,   making the conjecture of the existence
of a type I soliton of
the heterotic string all the
 more attractive.

\chapter {Compactified Type II String Theories}

Toroidally compactified type II superstring or supergravity theories have a
number of massless scalar
fields and their expectation values can all be regarded as coupling constants.
Any of these
can be used   to define a perturbation theory and one can ask which of the
$p$-branes are
perturbative states for a given coupling constant.
In this section we will consider this for the scalar field corresponding to the
dilaton in ten
dimensions, while the other coupling constants will be discussed in the next
section.
In each case, we will examine
the $p$-brane spectrum of the appropriate supergravity theory and look for the
perturbative states.
We will find that for the \lq stringy' coupling constant corresponding to the
dilaton in ten
dimensions, the
 perturbative spectrum at weak coupling is that of a string theory, while in
almost all other cases
(for strong stringy coupling or for other coupling constants) the perturbative
states
are those of a field theory. The main exception is in six dimensions, where the
perturbative
spectrum at strong string coupling is again that of a string theory.
This then supports string/string duality in $D=6$ dimensions, but not
string/$(D-5)$-brane duality
in $D>6$ dimensions.
More precisely, the stringy strong coupling limit in $D>6$ is some theory whose
perturbative states
are particles, not strings or $p$-branes.
It will have $p$-brane solitons, but is not a {\it perturbative } $p$-brane
theory involving the sum
over $p$-brane world-volume topologies etc.

In $D$ dimensions, a $(p+1)$-form gauge field couples to an electric $p$-brane
and a magnetic
$\tilde p$-brane, where
$\tilde p= D-p-4$.
 On toroidally compactifying to $D$ dimensions, the NS/NS 2-form gauge field
$b_{\mu \nu}$
gives rise to a D-dimensional 2-form, $d=10-D$ vector gauge fields and
$d(d-1)/2$ scalars, while
the metric gives rise to an additional $d$ vectors and $d(d+1)/2$ scalars. The
2-form couples to a
string and a
$(D-5)$-brane and the $2d$ vectors couple to
$2d$ {}
$0$-branes and
$d$ {} $(D-4)$-branes.
The
scalars from $b_{\mu \nu}$ only appear through their derivatives and so are
axion-like   in that there
is a classical symmetry under which they are shifted by constants.
Each of these couples to a $(D-3)$-brane.
There are also  RR $(p+1)$-form gauge fields for various values of $p$ coupling
to $p$-branes and
$\tilde p$-branes, and RR axionic scalars coupling to  $(D-3)$-branes (this can
be thought of as
the case $p=-1$, for which there is only a magnetic brane).

The mass formula (4.5) then gives the following brane scan for the NS-NS branes
($\gamma=2$):

\vskip 1cm
\vbox{
\begintable
 Value of $p$ | $1 $ |$ \tilde 1=D-5$  | $0$  | $\tilde 0=D-4$   |
${\tilde {-1}}=D-3$
\elt $\rho_p= \epsilon_p/p+1$ | 0 | 2/D-4 | 0 | 2/D-3 | 2/D-2 \elt
 $M_p/T_p$ in string metric | 1 | $g^{-2}$ | 1 | $g^{-2}$ | $g^{-2}$ \elt
  $M_p/T_p$ in general metric | $g^{-2\alpha}$ | $g^{- (2+(D-4)\alpha)}$ |
$g^{- \alpha}$ |
$g^{-(2+(D-3)\alpha)}$ | $g^{- (2+(D-2)\alpha)}$
\endtable

\centerline{{\bf Table  4} Type II NS-NS Brane-Scan in $D$ Dimensions.}}
 \vskip .5cm
\noindent while the RR branes ($\gamma=0$) have the following scan:

\vskip 1cm
\vbox{
\begintable
 {} | $p $  |$ \tilde p=D-p-4$
\elt
$\rho_p= \epsilon_p/p+1$ | $1/p+1$ | $1/D-p-3=1/\tilde p +1$ \elt
 $M_p/T_p$ in string metric | $g^{-1}$ | $g^{-1}$   \elt
  $M_p/T_p$ in general metric | $g^{- (1+(p+1)\alpha)}$ | $g^{- (1+(\tilde
p+1)\alpha)}$ =
$g^{- (1+(D-p-3)\alpha)}$
\endtable

\centerline{{\bf Table  5} Type II RR Brane-Scan in $D$ Dimensions.}}
 \vskip .5cm
\noindent The two columns give the electric and magnetic branes respectively
for $p\ge 0$, while for
the axionic branes with $p=-1$, there are no electric branes but there are
magnetic $(D-3)$-branes.

In all dimensions, the minimum value of $\rho $ is $0$ and is carried by the
NS-NS string and  $2d${}
$0$-branes.
By this it is meant that there are $2d$ \lq elementary' $p$-branes carrying
just one unit of charge
[\HT], while the full $0$-brane spectrum of charges is given by a
$2d$-dimensional lattice
preserved by the T-duality group $SO(d,d;\Z)$.
These then constitute the perturbative spectrum at weak coupling and are the
expected
result from a compactified string theory: a string plus $d$ winding modes and
$d$ momentum modes.

The maximum value of $\rho$ for $10>D> 6$ (and for type IIA in $D=10$) is $1$
and is carried by the
RR
$0$-branes so that the perturbative states of the strong coupling theory are
those of a field
theory, which as we shall see in the next section are those of $11$-dimensional
supergravity
compactified on a
$(d+1)$-torus with coupling constant corresponding to the inverse volume of the
torus, as expected
from [\Witten]. For $D\le 6$, however, the RR 0-branes no longer win over the
other branes.
In $D=6$, the maximum value of $\rho$   is again $1$ and is carried by both the
8 RR $0$-branes and
the magnetic NS-NS string resulting from a 5-brane wrapped around the 4-torus.
The perturbative spectrum at strong coupling is then again that of a
compactified string theory,
consisting of a 6-dimensional string plus 4 momentum modes and 4 winding modes.
This is the expected string-string duality.
In $D=5$, the maximum value of $\rho$ is $2$ and is carried by the NS-NS
magnetic $0$-brane
coming from wrapping the NS-NS 5-brane round the 5-torus. The strong-coupling
perturbative theory is
the field theory emerging in the large radius limit of the $D=6$ type II theory
compactified on a
circle.

Thus above six dimensions the RR 0-branes win out and the resulting
dynamics is governed by a field theory, but in less than six dimensions
 the strong coupling theory
is dominated by   magnetic NS-NS
branes. In the intermediate case of six dimensions, the RR 0-branes and the
NS-NS string have
equal values of $\rho$ and so
are both
perturbative.

\chapter {Compactified Type II Theories and U Duality}

Consider a supergravity theory in $D<10$ dimensions with at least half the
maximal number of
supersymmetries (e.g. $N \ge 4$ in $D=4$). The scalar fields take values in the
coset space
$G/G_c$ where $G$ is the supergravity duality group of
symmetries of the equations of motion and
$G_c$ is the maximal compact subgroup of $G$. The coupling constants are the
scalar expectation
values.  One of the lessons of U-duality is that the coupling constants are all
on the same footing and in principle any of them can be given the preferential
treatment usually
afforded to the dilaton in string theory.
A given  coupling constant $g$ is  $e^T$
where $T$ is a non-compact element of the Lie algebra of $G$ and is associated
with a scalar
expectation
$T=
{\langle \phi \rangle }$ for some $\phi \in  G/G_c$. We will be interested in
the dependence of
masses on the coupling
$g$ and in particular in
the strong coupling limit $g \to \infty$ and the weak coupling limit $g \to
0$. Our aim will be to find those states that are  in the perturbative spectrum
at weak coupling
and at strong coupling

The curve $e^{tT}$ parameterised by
$t$ defines an $\R ^+$ subgroup of $G$ and induces a maximal embedding
$$ H\times \R ^+ \subset G
\eqn\embed$$
for some  $H$.
Factoring both sides by their maximal compact subgroup induces a decomposition
of the moduli space of
the form
$$
{G \over G_c} \sim {H \over H_c} \times  \R ^+  \times \R^M$$
where $H_c$ is the maximal compact subgroup of $H$ and $\R^M$ is a vector space
equipped with a representation of $H$ [\AM]. A boundary of the moduli space is
then given by  going
to infinity in the $\R ^+$ direction, i.e. by taking
$t
\to
\infty$. It was argued in [\Witten,\AM] that it is sufficient to consider those
subgroups $H$ that can
be obtained by   removing from the Dynkin diagram of $G$ a single vertex
representing a non-compact
simple root.
In a quantum theory for which the supergravity is   a low-energy effective
action (as in string
theory), the classical supergravity duality group is broken to a discrete
subgroup $G(\Z)$, which
is the conjectured U-duality symmetry of the theory [\HT].

The $n$-form gauge fields of the supergravity theory formulated with respect to
the
Einstein metric transform as a representation
$R_n$ of
$G$ and this implies that the magnetic $p$-branes with $p=D-n-3$ that couple to
these have
charges transforming as the $R_n$ representation of $G$ while the corresponding
electric $p$-branes
with
$p=n-1$ have charges that transform according to the contragredient
representation $R_n'$.
In either case, the embedding
\embed\ induces a decomposition of the $G$ representation $R$:
$$
R \to \oplus _i (R_i^H)^ {\tilde \epsilon _i}
\eqn\rdeco$$
where the $R_i^H$ are representations of $H$ and the $\tilde \epsilon _i$ are
the
corresponding $\R^+$ weights. This implies that the mass per unit $p$-volume of
the $p$-branes
transforming as the $R_i^H$ of $H$ have the following $g$-dependence with
respect to the Einstein
metric $\gmn$:
$$
M_p = g^{-\tilde \epsilon _i} T_p
\eqn\gdepwt$$
where the effective tension $T_p$ can depend on the other coupling constants in
${H / H_c}    \times \R^M$.
Then for the general metric $e^{2\tilde \alpha \phi}\gmn$, the mass per unit
volume becomes
$$
M_p = g^{-(p+1)(\tilde\rho _i+\tilde \alpha)} T_p
\eqn\msss$$
where
$$\tilde\rho _i= {1 \over p+1}\tilde \epsilon _i
\eqn\tildrhis$$
The $p$-branes can then be assigned to representations $({\bf R^H_i})^{\tilde
\epsilon_i}_{\tilde
\rho_i}$ with $H$ representation ${\bf R^H_i}$ and $\R ^+$ weight $\tilde
\epsilon_i $, while
$\tilde\rho_i$ is defined by
      $\tilde\rho_i=\tilde \epsilon_i/(p+1)$.
In this way the parameters $\tilde\rho$ can be found for all $p$-branes and all
choices of $H$.
For a given $H$ the $p$-branes corresponding to the maximum and minimum values
of $\tilde\rho$
are the perturbative states of the strong and weak coupling limits,
respectively.

We shall now examine the consequences of this in a number of examples, finding
broad agreement with
the conjectures of [\Witten,\AM].
This gives an important check on these conjectures, which were based on the
0-brane spectrum only.
It is a non-trivial result that the behaviour of the $p$-brane spectrum
supports these conjectures,
rather than e.g. the conjectured string/(D-5)-brane duality. Furthermore, this
analysis
will enable us to treat cases that could not be analysed in
[\Witten,\AM].
 We shall consider first
 the
maximal supergravity theories in $D$ dimensions, given by compactifying
11-dimensional supergravity
on an
$11-D$ torus or 10-dimensional type IIA or IIB supergravity  a $10-D$ torus,
which arise from
compactified M-theory or compactified type II superstrings.

\section{Type II in D=7, Duality Group $G=SL(5)$}

The maximal supergravity in 7 dimensions has duality group $SL(5)$ with 5
2-form gauge fields
transforming as the ${\bf 5}$
of $SL(5)$ and 10 1-form gauge fields
transforming as the ${\bf 10 '}$
of $SL(5)$. The 2-forms couple to magnetic 2-branes transforming as a ${\bf 5}$
and electric 1-branes
transforming as a
${\bf 5'}$.
The 1-forms couple to magnetic 3-branes transforming as a ${\bf 10'}$
and electric 0-branes
transforming as a
${\bf 10}$. There are two choices of $H$, $SL(4)$ and $SL(3)\times SL(2)$.

\subsection{The $SL(4)\times \R^+$ Decomposition}

Here $SL(4) \sim SO(3,3)$ is the T-duality group of the type II string in
$D=7$, so that $g$ can be
associated with the
string coupling constant of the type II string compactified on $T^3$.
 The decomposition of the  relevant
$SL(5)$ representations under $SL(4)\times \R^+$ is
$$
{\bf 5} = {\bf 1 }^4 + {\bf 4 }^{-1}, \qquad
{\bf 10} = {\bf 4 }^3 + {\bf 6 }^{-2}
\eqn\tryre$$
where
the exponent is the $\R ^+$ weight $\tilde \epsilon$.
This gives the following brane-scan

\vskip 1cm
\vbox{
\begintable
0-branes | $ {\bf 6 }^{-2}_{-2} +{\bf 4 }^{3}_{3}$ \elt
1-branes | $ {\bf 1 }^{-4}_{-2} + {\bf 4' }^{1}_{1/2}$ \elt
2-branes | $ {\bf 1 }^{4}_{4/3} + {\bf 4 }^{-1}_{-1/3}$ \elt
3-branes | ${\bf 4 '}^{-3}_{-3/4} + {\bf 6' }^{2}_{1/2}$
\endtable

\centerline{{\bf Table 6} $SL(4)\times \R^+$ Decomposition of $D=7$ Type II
Brane-Scan.}}
%\centerline{}

%\catcode`\|=12

 \vskip .5cm
\noindent (Recall that the $p$-branes are assigned to representations $({\bf
R^H_i})^{\tilde
\epsilon_i}_{\tilde
\rho_i}$ with $H=SL(4)$ representation ${\bf R^H_i}$, $\R ^+$ weight $\tilde
\epsilon_i $
and value of   $\tilde\rho_i=\tilde \epsilon_i/(p+1)$.)
The minimum value of $\tilde\rho$ is $\tilde\rho_{min}=-2$  and is carried by
the
string and six 0-branes that transform as a singlet and a ${\bf 6'}$
respectively under the T-duality
group $SO(3,3)$.
These are the  perturbative states of the weakly coupled $D=7$ type II string
-- the
fundamental string of the $D=10$ type II   theory gives a string plus 3
momentum mode 0-branes and 3
winding mode 0-branes on compactification on $T^3$.
The coupling constant $g$ is the
 string  coupling and $\phi $ is   the dilaton.
Thus the correct treatment of the $D=7$ supergravity theory as a perturbation
theory in $g$
is as  a string theory, as expected.

 The maximum value of $\tilde\rho$ is $\tilde\rho_{min}= 3$  and is carried by
four 0-branes, fitting into 4 $D=7$ supergravity multiplets that transform as a
  ${\bf 4'}$  under
$SL(4)$. This gives the same set of perturbative states that emerged in section
4 from
11-dimensional  supergravity compactified on a 4-torus of volume $V$ in the
large  $V$ limit, where
the group $SL(4)$ acted as the mapping class group of the $T^4$.
This is consistent with the proposal   [\Witten] that the  theory that emerges
in the strong
coupling limit of the seven dimensional type II string theory
is an 11-dimensional theory compactified on $T^4$. Moreover, we learn that
the   perturbative theory that emerges is a particle theory, not a
supermembrane theory.

\subsection{The $SL(2)\times SL(3)\times\R^+$ Decomposition}

 The   relevant
$SL(5)$ representations decompose under $SL(2)\times SL(3)\times\R^+$ as
$$
{\bf 5} = ({\bf 2,1 })^3 + ({\bf 1,3 })^{-2}, \qquad
{\bf 10} = ({\bf 1,1 })^6 + ({\bf 1,3' })^{-4}+ ({\bf 2,3 })^1
\eqn\tryredf$$
This leads to the  following brane-scan
\vskip 1cm
\vbox{
\begintable
0-branes | $ ({\bf 1,1 })^{6} _{6}+ ({\bf 1,3' })^{-4}_{-4}+ ({\bf 2,3
})^{1}_{1}$ \elt
1-branes | $ ({\bf 2',1 })^{-3}_{-3/2} + ({\bf 1,3' })^{2}_{1}$ \elt
2-branes | $ ({\bf 2,1 })^{3} _{1}+ ({\bf 1,3 })^{-2}_{-2/3}$ \elt
3-branes | $({\bf 1,1 })^{-6}_{-3/2} + ({\bf 1,3 })^{ 4}_{1}+ ({\bf 2',3'
})^{-1}_{-1/4}$
\endtable

\centerline{{\bf Table 7} $SL(2)\times SL(3)\times\R^+$ Decomposition of $D=7$
Type II  Brane-Scan.}
}

The minimum value of $\tilde \rho$ is $-4$ and is carried by 0-branes in the
$({\bf 1,3' })$
representation of $SL(2)\times SL(3)$ and is the same perturbative spectrum
that arises from type IIB {\it supergravity}
compactified on $T^3$ in the limit of large torus volume,
 with $SL(2)$ the type IIB duality group and
$SL(3)$ the $T^3$ mapping class group.
If we consider the type IIB {\it superstring}
compactified on $T^3$, the low energy effective effective action is the seven
dimensional
supergravity theory under consideration, and the spectrum of $p$-branes is the
same.
Thus this perturbation theory of the type IIB superstring gives a field theory,
not a string theory.

The maximum value of $\tilde \rho$ is $6$ and is carried by 0-branes in the
$({\bf 1,1 })$
representation of $SL(2)\times SL(3)$ and is the same perturbative spectrum
that arises from 8-dimensional {  supergravity} or $D=8$ type II superstring
compactified on a circle in the  large radius limit.

These results are in agreement with [\Witten], but clarify the nature of the
compactified theory.
Different perturbative  theories   emerge with different choices of
perturbation parameter,
corresponding to different choices of $H$.
Choosing $H=SL(4)$ as in the previous subsection gives a perturbative string
theory
in which the perturbation parameter is the string coupling,  related to the
dilaton in the
usual way, while
choosing $H=SL(2)\times SL(3)$  corresponds to perturbing in the inverse volume
of the torus, and
the perturbative theory that emerges is a field theory, not a string theory.

%%%%%%%%%

\section{Type II in D=6, Duality Group $G=SO(5,5)$}

The maximal supergravity in 7 dimensions has duality group $SO(5,5)$ with
16 1-form gauge fields
transforming as the spinor ${\bf 16}$
of $SO(5,5)$ and
 5 2-form gauge fields. The 5 3-form field strengths $H^a$, together with the
dual field
strengths\foot{The dual field
strength is given by the Hodge dual of the variation of the action with respect
to
$H^a$,  $\tilde H_a = *(\delta S/\delta H^a)$ transforming as the ${\bf 5'}$.
Thus the Bianchi
identities are
$dH^a=0$ and the field equations are $d \tilde H_a=0$.} form a 10-vector
transforming as a {\bf 10}
of  of
$SO(5,5)$. There are then 16 0-branes transforming as a ${\bf 16'}$, 16
2-branes transforming as a
${\bf 16}$, and 10 strings transforming as a {\bf 10}.
There are a number of choices of $H$ to consider.

\subsection{The $SO(4,4)\times \R^+$ Decomposition}

Here $SO(4,4)$ is the T-duality group of the type II string in $D=6$, so that
$g$ can be
associated with the
string coupling constant of the type II string compactified on $T^3$.
 Under $SO(4,4)\times \R^+$,
$$
{\bf 16} = {\bf 8_s }^1 + {\bf 8_c }^{-1}, \qquad
{\bf 10} = {\bf 1 }^2 + {\bf 1 }^{-2}+{\bf 8_v }^{0}
\eqn\trre$$
This gives the following brane-scan

\vskip 1cm
\vbox{
\begintable
0-branes | $ {\bf 8_s }^1_1 + {\bf 8_c }^{-1}_{-1}$ \elt
1-branes | $ {\bf 1 }^2_1 + {\bf 1 }^{-2}_{-1}+{\bf 8_v }^{0}_0$ \elt
2-branes | $ {\bf 8_c }^1_{1/3} + {\bf 8_s }^{-1}_{-1/3}$
\endtable

\centerline{{\bf Table 8} $SO(4,4)\times \R^+$ Decomposition of $D=6$ Type II
Brane-Scan.}
}

%\catcode`\|=12

 \vskip .5cm

The maximum value of $\tilde\rho$ is $\tilde\rho_{max}=1$  and is carried by
the
string and 8 0-branes that transform as a singlet and an ${\bf 8}$ respectively
under the T-duality
group SO(4,4).
These are the  perturbative states of the weakly coupled $D=6$ type II string
-- the
fundamental string of the $D=10$ type II   theory gives a string plus 4
momentum mode 0-branes and
4 winding mode 0-branes on compactification on $T^4$.
 The minimum value of $\tilde\rho$ is $\tilde\rho_{min}=-1$  and is again
carried by   a
string and 8 0-branes, reflecting the self-duality of the theory that is
predicted by U-duality.
Both strong and weak coupling limits give the same perturbative string theory,
which is that of a
compactified type II string.

\subsection{The $SL(5)\times \R^+$ Decomposition}

 The decomposition of the  relevant
 representations under $SL(5)\times \R^+$  is
$$
{\bf 16} = {\bf 1 }^{-5} + {\bf 5' }^{3}+{\bf 10}^{-1}, \qquad
{\bf 10} = {\bf 5 }^2 + {\bf 5' }^{-2}
\eqn\tryrefds$$
The resulting brane-scan is

\vskip 1cm
\vbox{
\begintable
0-branes | $ {\bf 1 }^{-5}_{-5} + {\bf 5' }^{3}_3+{\bf 10}^{-1}_{-1}$ \elt
1-branes | $ {\bf 5 }^2_1 + {\bf 5' }^{-2}_{-1}$ \elt
2-branes | $ {\bf 1 }^{5}_{5/3} + {\bf 5 }^{-3}_{-1}+{\bf 10'}^{1}_{1/3}$
\endtable

\centerline{{\bf Table 9} $SL(5)\times \R^+$ Decomposition of $D=6$ Type II
Brane-Scan.}
}
%\catcode`\|=12

 \vskip .5cm

The maximum value of $\tilde \rho$ is $3$ and is carried by 5  0-branes,
fitting into 5 $D=6$
supergravity multiplets that transform as a   ${\bf 5'}$  under
$SL(5)$. This gives the same set of perturbative states that emerged in section
4 from
11-dimensional  supergravity compactified on a 5-torus of volume $V$ in the
large  $V$ limit, where
the group $SL(5)$ acted as the mapping class group of the $T^5$.

The minimum value of
$\tilde \rho$ is $-5$ and is carried by the $SL(5)$ singlet  0-brane,
and the perturbative theory is that of 7-dimensional type II supergravity
compactified on a
circle, in the large radius limit.

%%%%%%%%%%%%%%%%%%%%%%%%%%%%%%%%%%%%%%%%%%%%%%%%%%%%%%%%%%%

\subsection{The $SL(4)\times SL(2)\times\R^+$ Decomposition}

 The   relevant
$SO(5,5)$ representations decompose under $SL(4)\times SL(2)\times\R^+$ as
$$\eqalign{
{\bf 16} &= ({\bf 4,1})^1 + ({\bf 4,1 })^{-1}+ ({\bf 4',2 })^{0},
\cr
{\bf 10} &= ({\bf 1,2 })^1 + ({\bf 1,2 })^{-1}+ ({\bf 6,1})^0
\cr}
\eqn\tdf$$
Both the
  maximum and minimum values of $\tilde \rho$ ($\pm 1$) are carried by 0-branes
in the $({\bf 4,1
})$ representation of $SL(4)\times SL(2)$, so that the theory is self-dual
 and the same perturbative theory as arises in both these limits also emerges
from
   the type IIB theory
compactified on $T^4$ in the limit of large or small torus volume,
 with $SL(2)$ the type IIB duality group and
$SL(4)$ the $T^4$ mapping class group.

%%%%%%%%%%%%%%%%%%%%%%%%%%%%%%%%%%%%%%%%%%%%%%%%%%%%%%%%%%%
%%%%%%%%%%%%%%%%%%%%%%%%%%%%%%%%%%%%%%%%%%%%%%%%%%%%%%%%%%%

\subsection{The $SL(2)\times SL(2)\times SL(3)\times\R^+$ Decomposition}

 The   relevant
$SO(5,5)$ representations decompose under $SL(2)\times SL(2)\times
SL(3)\times\R^+$  as
$$
\eqalign{
{\bf 16} & = ({\bf 2,1,1})^3 + ({\bf 2,1,3 })^{-1}+ ({\bf 1,2,1 })^{-3}+ ({\bf
1,2,3' })^{1},
\cr
{\bf 10} &= ({\bf 2,2,1 })^0 + ({\bf 1,1,3})^{2}+ ({\bf 1,1,3 })^{-2}
\cr}
\eqn\yredf$$
The maximum and minimum values of $\tilde \rho$ ($\pm 3$) are carried by
0-branes in the $({\bf
2,1,1 })$ and $({\bf
1,2,1 })$ representations of $SL(2)\times SL(2)\times SL(3) $
and these correspond to the
$D=8$ theory compactified on $T^2$ in the limits of large volume and large
values of the complex
structure modulus, respectively.

%%%%%%%%%%%%%%%%%%%%%%%%%%%%%%%%%%%%%%%%%%%%%%%%%%%%%%%%%%%%%%%%%%%%

\section{Type II in D=5, Duality Group $G=E_6$}

The maximal supergravity in 6 dimensions has duality group $E_{6(6)}$ with
27 1-form gauge fields
transforming as the  ${\bf 27}$
of $E_6$. These couple to 27 electric 0-branes and 27 magnetic strings,
transforming
as a ${\bf 27'}$ and ${\bf 27}$ respectively.

%%%%%%%%%%%%%%%
\subsection{The $SO(5,5)\times\R^+$ Decomposition}

 The   ${\bf 27}$ decomposes under $SO(5,5)\times\R^+$  as
$$
{\bf 27} = {\bf  1}^ 4+ {\bf 10 }^{-2} +{\bf  16}^1
\eqn\yredf$$
The  minimum value of $\tilde \rho$ is $ -2$ and is   carried by  a string and
10 0-branes in the
$({\bf 10 })$ of the T-duality group $SO(5,5)$, corresponding to the weakly
coupled  $D=5$ type II
string theory.
The  maximum value of $\tilde \rho$ is $ 4$ and is   carried by  a single
0-brane,
corresponding to the large radius limit of the $D=6$ type II theory
compactified on a circle.

%%%%%%%%%%%%%%%%%
%%%%%%%%%%%%%%%
\subsection{The $SL(6)\times\R^+$ Decomposition}

 The   ${\bf 27}$ decomposes under $SL(6)\times\R^+$  as
$$
{\bf 27} = {\bf  6'}^1 + {\bf 6' }^{-1} +{\bf 15 }^ 0
\eqn\yref$$

The maximum and minimum values of $\tilde \rho$ ($\pm 1$) are each carried by 6
0-branes in the
${\bf  6'}$
 representation  of $SL(6) $ so that the theory is self-dual in this direction
in coupling constant
space and either limit corresponds to the
$D=11$ theory compactified on $T^6$ in the limits of large or small torus
volume.

%%%%%%%%%%%%%%%%%%%%%%%%%%%%%%%%

\subsection{The $SL(2)\times SL(5)\times\R^+$ Decomposition}

 The   ${\bf 27}$ decomposes under $SL(2)\times SL(5)\times\R^+$  as
$$
{\bf 27} = ({\bf 2,1 })^{5} + ({\bf 2,5'  })^{-1} +({\bf 1,5 })^{-4} +({\bf
1,10 })^{2}
\eqn\redf$$

The maximum  value of $\tilde \rho$ is $5$ and is    carried by 0-branes in the
$({\bf
2,1 })$   representation  of $SL(2)\times SL(5) $
and   corresponds to the
$D=7$ theory compactified on $T^2$ in the limit of large torus volume.
The
  minimum value is $\tilde \rho=-4$ and is    carried by 0-branes in the $({\bf
1,5 })$   representation  of $SL(2)\times SL(5) $
and   corresponds to the
$D=10$ type IIB theory compactified on $T^5$.
%%%%%%%%%%%%%%%%%%%%%%%%%%%%%%%%
%%%%%%%%%%%%%%%%%%%%%%%%%%%%%%%%
\subsection{The $SL(3)\times SL(3)\times SL(2)\times\R^+$ Decomposition}

 The   ${\bf 27}$ decomposes under $SL(2)\times SL(5)\times\R^+$  as
$$
{\bf 27} = ({\bf 3,3',1 })^{0} +
 ({\bf  3,1,1})^{-2} +({\bf  3,1,2})^{1} +({\bf 1,3',1 })^{2} +({\bf
1,3',2'})^{-1}
\eqn\yro$$

The maximum and minimum values of $\tilde \rho$ ($\pm 2$) are carried by
0-branes in the $({\bf
3,1,1 })$ and $({\bf
1,3',1 })$ representations
and these correspond to the
$D=8$ theory compactified on $T^3$.

%%%%%%%%%%%%%%%%%%%%%%%%%%%%%%%%
%%%%%%%%%%%%%%%%%%%%%%%%%%%%%%%%
\chapter {Heterotic Theories and $K_3$ Compactifications}

In the previous sections we have considered theories with maximal supersymmetry
whose low energy
supergravity effective action comes from toroidally compactifying $N=2$
supergravity theories from
ten dimensions. In this section, we will turn to theories with half the maximal
number of
 supersymmetries -- $N=1$ in $D=10$ or $N=4$ in $D=4$ -- which can result from
toroidal
compactification of heterotic or type I theories, or from compactification of
type II theories on
$K_3 \times T^n$. In each case, the brane-scan is again determined by
low-energy supersymmetry,
so we shall start from the brane-scan of the supergravity theory, find which
are the perturbative
states for the various coupling constants, and then seek an interpretation of
the
resulting perturbation theory.

In $D>4$ dimensions, the  half-maximal supergravity theory coupled to $n$
abelian  vector
multiplets (resulting from toroidal compactification of $N=1$ theories in
$D=10$) has scalar fields
lying in the coset space
$$
{O(d,d+m)\over O(d) \times O(d+m) }\times \R ^+
\eqn\scal
$$
where $d=10-D$ and $m=n-d$
and has $2d+m$ vector fields transforming as the vector representation of the
duality group
$O(d,d+m)$. For the heterotic string, $m=16$, $O(d,d+16)$ is the T-duality
group and the dilaton
takes values in the
$\R ^+$ factor in \scal.
The 2-form  gauge field
couples to a string and a $(D-5)$-brane, while the vector fields couple to
 {} $0$-branes whose charges take values in a $2d+m$-dimensional lattice and
{}$(D-4)$-branes
whose charges take values in the $2d+m$-dimensional dual lattice. In
$D
\le 8$ dimensions, the  ten-dimensional 2-form gives rise to
$d(d-1)/2$ scalars that couple to magnetic $(D-3)$-branes.
As described in section 5.3, there may also be other branes, possibly related
to type I strings.
However, in $D\le 7$ the branes considered are precisely those that come from
compactifying 11-dimensional supergravity or M-theory on $K_3 \times T^N$
(where $N=7-D$) and    as
yet no
  extra exotic branes are suspected in these theories.
 We shall start by studying the known branes and investigating their
dependence on the various coupling constants in \scal.

\section {The Heterotic Dilaton}

In this section, we shall investigate perturbation theory in the coupling
constant $g$ in the $\R
^+$ factor of \scal, which corresponds to the heterotic string dilaton $\Phi$,
$g=e^{\langle \Phi
\rangle }$. The   brane scan is

\vskip 1cm
\vbox{
\begintable
 Value of $p$ | $1 $ |$ \tilde 1=D-5$  | $0$  | $\tilde 0=D-4$
\elt
$O(d,d+m)$ Representation | {\bf 1}|{\bf 1}| {\bf$2d+m$} |{\bf$2d+m$ }  \elt
 $\rho_p= \epsilon_p/p+1$ | 0 | 2/D-4 | 0 | 2/D-3  \elt
 $M_p/T_p$ in string metric | 1 | $g^{-2}$ | 1 | $g^{-2}$   \elt
  $M_p/T_p$ in general metric | $g^{-2\alpha}$ | $g^{- (2+(D-4)\alpha)}$ |
$g^{- \alpha}$ |
$g^{-(2+(D-3)\alpha)}$
\endtable

\centerline{{\bf Table 10} Heterotic Brane-Scan in $D$ Dimensions.}}
 \vskip .5cm
\noindent  In addition there
are $(D-3)$-branes that couple to certain scalars, such as those that arise
from the 2-form gauge field on compactification and for which  $\rho_p=2/D-2$.
The perturbative states at weak coupling $(\rho=0)$
are, as expected, a string plus
$2d+m$ 0-branes (generating a $2d+m$ dimensional lattice of supermultiplets).
The latter
consist of
$d$ momentum and
$d$ winding modes from a string on a
$d$-torus, plus the $m$ heterotic winding/momentum modes from a left-moving
string on an $m$-torus;
for the heterotic string,
$m=16$. Of these $2d+m$ 0-branes and 6-branes, only $d$ 0-branes and $d$
6-branes saturate a
Bogomolnyi bound and preserve half the supersymmetry; the others are not
supersymmetric. For strong
coupling, the maximum value of
$\rho$ in the table is
$2/D-4$ and is carried by the ($D-5$)-brane. If there are no other branes in
the spectrum with
higher values of
$\rho$, then this would imply that the strong-coupling perturbative spectrum
consists of a
($D-5$)-brane.  In section 5.3, however, reasons were given for expecting new
type I  branes with a
higher value of
$\rho$ in
$D=10$, and the conjectured type-I/heterotic duality in $D\ge 8$ dimensions
[\Witten] suggests
that the highest value of $\rho$  should   be carried by such branes in
$D=8,9,10$.
We shall therefore concentrate on $D<8$ dimensions here.

In $D=6$, the strong coupling perturbative spectrum consists of the
($D-5$)-brane, which in this case
is a string. There are   no 0-branes, so this doesn't correspond to the
spectrum of a toroidally
compactified heterotic string, but is precisely the weak-coupling perturbative
spectrum of   the type
IIA  string theory compactified on $K_3$, with type IIA coupling constant
$g'=1/g$.
Thus the strongly coupled heterotic string in $D=6$ is described
by a weakly coupled type IIA string theory compactified on $K_3$, as expected
[\HT,\Witten].

Consider an 11-dimensional theory (supergravity or M-theory) compactified on
$K_3\times S^1$.
Large and small values of $g$ correspond to large and small values of
the radius of the circle, respectively. Thus, we learn that in the large radius
limit, the
perturbation theory
arising from expanding in the inverse radius is a string theory, as expected
from the corresponding
results for compactifying from 11 dimensions to 10 on $S^1$.

In $D=5$ dimensions, the strong coupling spectrum consists of a single 0-brane
supermultiplet,
and this is described by the supergravity theory resulting from type IIB
compactified on $K_3\times
S^1$, with the $g\to \infty$ limit  corresponding to that in which the radius
of the circle becomes
infinite [\Witten].

In $D=7$ dimensions, however, the
BPS-saturated brane with the highest value of $\rho$ is a supermembrane.
Precisely the same brane-scan arises from 11-dimensional supergravity
compactified
on a $K_3$ with radius $g^{1/3}$, and it was suggested in [\Witten] that the
strongly coupled $D=7$
heterotic string should correspond to $K_3$-compactified 11-dimensional
supergravity in the large
radius limit. However, if this were the case, then one would expect the
perturbative spectrum to
consist of 0-branes instead of membranes; this was certainly the way things
worked so far.
There seem to be two possibilities. The first is that  this is indeed the
correct perturbative
spectrum and that the large
$g$ behaviour is described  by a perturbative supermembrane theory in $D=7$, as
considered in
[\PKTb]; it is remarkable that in contrast to the case of 11-dimensions there
is a coupling constant
and such a theory could exist.
The second possibility is that there are other
non-perturbative states
  in the spectrum  that govern the strong coupling, perhaps corresponding to
branes
 with higher values of $\rho$.
This would be analogous to what happened in the type IIA theories: the extra
states needed to \lq beat' the ($D-5$)-branes in $D\ge 7$
arose from the RR sector and  the strong coupling theory was described in terms
of RR 0-branes.
This possibility seems the more likely, and might work as follows.\foot{I would
like to thank John Schwarz for this suggestion.} 11-dimensional supergravity
compactified
on   $K_3$ has  Kaluza-Klein scalars in its spectrum that become massless in
the large volume limit.
They are not BPS-saturated and so, although they should be in the pertrurbative
spectrum at large volume, they should not be expected to persist for other
values of the coupling.
With the choice $\alpha=-1$, these Kaluza-Klein $0$-branes have $M/T \sim 1$,
while the
string, membrane, heterotic 0-branes and 3-branes have $M/T$ given by
$g^2,g,g,g^2$ respectively.
Thus if the  heterotic string in $D=7$ has such non-BPS 0-brane states that are
metastable at strong coupling but which may not continue to states at weak
coupling, then these would be the perturbative states at strong coupling and
this is what must happen if the  strong coupling limit is to be
$K_3$-compactified 11-dimensional supergravity.

\section{Other Coupling Constants}

We now turn to the
coupling constants in the $O(d,d+m)/O(d)\times O(d+m)$ factor in \scal.
{}From [\AM], the distinct possibilities are labelled by an integer $n$ and
constitute going to a
boundary in the moduli space in which
$$
{O(d,d+m)\over O(d) \times O(d+m) }  \to
{O(d-n,d-n+m)\over O(d-n) \times O(d-n+m) }\times {SL(n,\R) \over SO(n)} \times
\R ^+ \times \R
^M
\eqn\bound
$$
for some $M$. We will be interested in the dependence on the coupling constant
$g$ corresponding to
the
$\R^+$ factor in \bound.
The 2-form is neutral under $O(d,d+m)$ and so the corresponding string and
$(D-5)$-form will not
couple to this coupling constant. The vector fields transform as the
fundamental representation,
which decomposes under
$$
{O(d,d+m) }  \to
{O(d-n,d-n+m)  }\times {SL(n,\R)  } \times \R ^+
\eqn\bounda
$$
as
$${\bf 2d+m} \to ( {\bf 2(d-n)+m, 1} )^0 + ( {\bf1,n})^1 +({\bf1,n'})^{-1}$$
The resulting brane-scan for the $0,1$ branes and their magnetic duals is
\vskip 1cm
\vbox{
\begintable
0-branes | ($ 2(d-n)+m, 1 )^0_0 + ( 1,n)^1_1 +(1,n')^{-1}_{-1}$ \elt
1-branes | $1^0_0$
\elt
(D-5)-branes | $1^0_0$
\elt
(D-4)-branes | $ (  2(d-n)+m, 1 )^0_0 + ( 1,n')^1_{1/D-3} +(1,n)^{-1}_{1/D-3}$
%\elt
%(D-3)-branes | $1^0_0$
\endtable

\centerline{{\bf Table 11} Heterotic  Brane-Scan.}}
 \vskip .5cm
In $D\ge 4$ the maximum and minimum values of $\rho$ are $\pm 1$ and are
carried by $n$ 0-branes, so
that the perturbative spectrum for both $g$ and $1/g$ consists of $n${}
$0$-brane multiplets in
$D>4$ and $2n${} $0$-brane multiplets in $D=4$.

We now turn to the interpretation of these results, using the
discussion of [\AM]. The factor ${SL(n,R) \over SO(n)}$ in \bound\ is the
moduli space of
fixed-volume metrics on an
$n$-torus. From earlier discussions, the perturbative spectrum of a
supergravity theory
compactified on $T^n$ with respect to the coupling constant representing the
volume of $T^n$
consists of $n$ 0-brane multiplets in the lower dimension, and this is
consistent with the
results just found. The factor
$${O(d-n,d-n+m)\over O(d-n) \times O(d-n+m) }$$ in \bound\ can be interpreted
as a heterotic
string moduli space  or as the moduli space of
$K_3$ or some squashed $K_3$
for certain values of $d,n,m$. First,
$${O(M,M+16) \over O(M)\times O(M+16)}$$ is the moduli space of heterotic
strings in
$10-M$ dimensions. Next $${O(3,19) \over O(3)\times O(19)}$$ is the Teichm\"
uller space of
Ricci-flat metrics of fixed volume on $K_3$ while $${O(4,20) \over O(4)\times
O(20)}$$ is the moduli
space of conformal field theories on $K_3$ [\AMK]. The spaces $${O(3-n,19-n)
\over O(3-n)\times
O(19-n)}$$ for
$n=1,2,3$ were identified in [\AM] as the moduli spaces of certain
degenerations of $K_3$ down to
$4-n$ dimensions. It will be convenient to denote the $m=4-n$ dimensional
squashed $K_3$ as
$\Xi^m$, with $\Xi^4=K_3$. There are in fact two distinct 1-dimensional
degenerations [\AM], which we
will denote $\Xi^1_1$ and $\Xi^1_2$.
It was argued in [\AM] that taking the strong coupling limit in certain
directions in coupling
constant space corresponds to going to boundaries of the $K_3$ moduli space in
which $K_3$
degenerates to
one of the $\Xi^m$. As in [\AM], it will be assumed here that taking this limit
makes sense and
corresponds to a compactification of M-theory    on the $\Xi^m$.

\subsection {$D=6$}

Consider   the case of the six-dimensional theory with scalar coset space
$$
{O(4,20)\over O(4) \times O(20) }   \times \R ^+
\eqn\six
$$
which arises from the toroidally compactified heterotic string, type IIA
compactified on $K_3$ and
the 11-dimensional theory compactified on $K_3\times S^1$.
For the coupling constant in the $\R^+$ factor, the weak coupling theory is the
perturbative
heterotic string and the strong coupling is described by the perturbative type
IIA string, as
discussed in section 8.1.

 Consider first the degeneration
$$
{O(4,20)\over O(4) \times O(20) }   \to
{O(3,19)\over O(3) \times O(19) }\times     \R ^+  \times \R
^{22}
\eqn\boundy
$$
The charged perturbative spectrum in $g$ at weak coupling  and in $1/g$ at
strong coupling both
consist of a single six-dimensional {}  $0$-brane supermultiplet, so that both
the weak and strong
coupling limits are field theories.
 The limit $g \to \infty$ for the coupling constant $g$ corresponding to  the
$\R^+$ factor in
\boundy\  corresponds, for the heterotic string, to a boundary of the
$T^4$  moduli space in which one of the circles in
$T^4$ becomes large, so that in the limit the theory decompactifies to the
7-dimensional heterotic
string, which has moduli space $${O(3,19)\over O(3) \times O(19) }\times     \R
^+$$
The limit $g \to 0$ corresponds to one of the circles becoming small, but this
is related by
T-duality to the large radius limit of the same theory, so that the theory is
self-dual and the weak and strong coupling limits (with respect to $g$) define
equivalent theories.
In the limit $g \to \infty$ (or $g \to 0$), one recovers 7-dimensional Lorentz
invariance.

Consider now the interpretation of these limits in terms of the 11-dimensional
M  theory.
The  limits $g \to 0$ and $ g \to \infty$ correspond  to   boundaries of the
moduli space of
compactifications of the 11-dimensional theory on $K_3\times S^1$ which lead to
  7-dimensional theories
and so must be  the ones in which the circle becomes large or small.
The large radius limit decompactifies to the 7-dimensional theory resulting
from compactifying from
11-dimensions on $K_3$, which is expected to correspond to the 7-dimensional
heterotic string
[\Witten].
We also learn that the small radius limit is equivalent to the large radius
limit (using the
heterotic equivalence), so that the 11-dimensional M-theory must also exhibit
some form of
T-duality. Thus the perturbation theory of the $D=6$  heterotic string
with respect to the coupling constant that vanishes when the theory
decompactifies to $D=7$
is described by the field theory arising from compactifying the 11-dimensional
theory on $K_3\times
S^1$ and expanding in the inverse circle radius, as expected from [\Witten].

Similar remarks apply to other degenerations.
For
$$
{O(4,20)\over O(4) \times O(20) }  \to
{O(2,18)\over O(2) \times O(18) }\times { SL(2,\R) \over U(1)}  \times \R ^+
\times \R
^{41}
\eqn\boundyb
$$
we obtain two 0-brane supermultiplets (transforming as a {\bf 2} of $SL(2,\R)$)
as the perturbative spectrum both at weak and strong coupling for the coupling
constant corresponding
to the $\R^+$ factor in \boundyb. This corresponds to (i) the 8-dimensional
heterotic string on
$T^2$ and (ii) the 11-dimensional theory on $\Xi^3\times T^2$, assuming this
limit makes sense
[\AM].
 In each case $g$ is related to the volume of the $T^2$, so that $g \to \infty$
corresponds to
decompactification to 8-dimensions, while the $g \to 0$ limit
defines an equivalent theory, related by T-duality.
 For
$$
{O(4,20)\over O(4) \times O(20) }  \to
{O(1,17)\over   O(17) }\times { SL(3,\R) \over SO(3)}  \times \R ^+ \times \R
^{57}
\eqn\boundyc
$$
we obtain three 0-brane supermultiplets
as the perturbative spectrum both at weak and strong coupling. This corresponds
to (i) the
9-dimensional heterotic string on
$T^3$   and (ii) the 11-dimensional theory on $\Xi^2\times T^3$.
Finally, for
$$
{O(4,20)\over O(4) \times O(20) }  \to
  { SL(4,\R) \over SO(4)}  \times \R ^+ \times \R
^{70}
\eqn\boundyd
$$
we obtain four 0-brane supermultiplets
as the perturbative spectrum both at weak and strong coupling. This corresponds
to (i) the
10-dimensional heterotic string on
$T^4$  (ii) the 11-dimensional theory on $\Xi^1\times T^4$.

Similar results apply in other dimensions $D\le 6$ (cf. [\AM]) with a
perturbative field theory
emerging in each case when expanding in couplings corresponding to scalars
other than the heterotic
dilaton. The results are consistent with the conjecture that the heterotic
string in
$D$ dimensions is equivalent to 11-dimensional M-theory compactified on $\Xi
^{11-D}$ and
 to the type IIA string compactified on
$\Xi ^{10-D}$, with $\Xi^n$ defined to be
$K_3\times T^{n-4}$ for $ n \ge 4$. This gives the following table.
\vskip 1cm
\vbox{
\begintable
 Dimension $D$| Symmetry Group G | M-Theory Compactified on:   \elt
 10 | $O(16)$ | $\Xi ^1  $   \elt
 9 | $O(1,17)$ | $\Xi ^2  $   \elt
 8 | $O(2,18)$ | $\Xi ^3  $  \elt
 7 | $O(3,19)$ | $K_3 $   \elt
 6 | $O(4,20)$ | $K_3\times T^1$   \elt
 5 | $O(5,21)$ | $K_3\times T^2$   \elt
 4 | $O(6,22)$ | $K_3\times T^3$
\endtable

\centerline{{\bf Table 12} Conjectured duals of $D$-dimensional heterotic
string from }
\centerline{ compactifying M-theory.}}
 \vskip .5cm
\noindent
There is considerable evidence supporting the  relations involving
 $K_3$ or $K_3\times T^n$ [\HT,\Witten,\AM,\HTE] but those involving the
degenerations of $K_3$
[\AM] are more speculative and deserve further investigation.

It is interesting to speculate on the relation of these conjectures to those of
Horava and Witten
[\Mwit] that the $E_8 \times E_8$ heterotic string in $D=10$ is related to
M-theory compactified from
11 dimensions on the orbifold $S^1/\Z_2$. Compactifying to $D=7$ on $T^3$
(without Wilson lines)
gives the $D=7$ heterotic string with gauge group $E_8 \times E_8\times
U(1)^6$, or M-theory
compactified  on $T^3 \times S^1/\Z_2$. However, the same theory should emerge
from the
11-dimensional theory compactified on an orbifold limit of $K_3$ at which
$E_8 \times E_8$ gauge
symmetry emerges [\Witten,\HTE]; it would be interesting to understand further
why
M-theory compactified on these two orbifolds should give the same theory. This
in turn might
shed some light on the construction of the $K_3$ degenerations $\Xi^m$:
M-theory on $\Xi^m$ should,
at an enhanced symmetry point of the $\Xi^m$ moduli space, be equivalent to
M-theory on $T^{m-1}
\times S^1/\Z_2$.

\chapter {The Chiral Theory in Six Dimensions}

The N=2 supersymmetric theories in six dimensions considered so far have had
(1,1) supersymmetry,
i.e. the two supersymmetries have opposite chirality.
There are also (2,0) theories in $D=6$ with both supersymmetries having the
same chirality.
For (2,0) supergravity coupled to $n$ (2,0) anti-symmetric tensor multiplets,
the scalar fields
take values in the coset space
$${O(5,n)
\over O(5)\times O(n)}
\eqn\mosd$$
There are $5+n$ anti-symmetric tensor gauge fields
$b_{\mu \nu }^I$ ($I=1, \dots , 5+n$) with corresponding field strengths
$H^I_{\mu \nu \rho}$ that
transform as a vector of $O(5,n)$. The scalars can be represented as a $(5+n)
\times (5+n)$ matrix
${\cal V}_{IJ}$ transforming under rigid $O(5,n)$ from the right and local
$O(5) \times O(n)$ from
the left.  Decomposing the
 $O(5) \times O(n)$ index  into an $O(5) $ index $a$ and an $  O(n)$ index $m$,
one can form the
$O(5,n)$-invariant
 field strength 3-forms
$$ G_a =  {\cal V}_{aI} H^I , \qquad K_m= {\cal V}_{mI}H^I
$$
The 5 field strengths $G_a$
  are   self-dual while the $n$ field strengths  $K_m$ are
anti-self-dual:
$$
G_a=*G_a, \qquad H_m=-* H_m$$
Decomposing the $O(n)$ index $m$ into an $O(5)$ index $a$ and an $O(n-5)$ index
$u$, $H_m \to
(H_a,H_u)$
and defining
$$ G^\pm _a = G_a \pm H_a
\eqn\defg$$
we obtain 5 unconstrained field strengths $G_a^+$, their duals $G_a^-=*G_a^+$,
and $n-5$
anti-self-dual field strengths $H_u$.

 The theory with $n=19$ emerges as the low-energy effective field theory for
the type IIB
superstring compactified on $K_3$, with U-duality group $O(5,21;\Z)$ [\AM].
The $p$-brane spectrum for general $n$ consists of 5 self-dual strings and $n$
anti-self-dual ones,
or equivalently 5 ordinary (non-chiral) strings and $n-5$ anti-self-dual ones.

Consider the compactification of this theory to $D=5$.
If the type IIB theory is compactified on $K_3\times S^1$ with circle of radius
$R$, it is
equivalent to the type IIA theory   compactified on $K_3\times S^1$ with circle
of radius $1/R$,
and this is in turn related to the $D=5$ heterotic string [\Witten] and the
11-dimensional theory
on $K_3\times T^2$. However, in $D=6$ the chiral supergravity theory is not
related to any of the
theories in   table 12.

The general decomposition is
$${O(5,21)
\over O(5)\times O(21)} \to  {O(5-n,21-n)
\over O(5-n)\times O(21-n)}\times {SL(n,\R)\over O(n)}\times \R^+ \times \R^{M}
 \eqn\mos $$
where $M=-{3\over 2}n^2+26n-{1\over 2} $.
The {\bf 26} representation of
$O(5,21)$ decomposes under
$$O(5,21) \to O(5-n,21-n) \times SL(n,\R) \times \R^+$$
 as
$${\bf 26} \to ({\bf 26-2n}, {\bf 1})^0 +({\bf 1},{\bf n})^1 +({\bf 1},{\bf
n'})^{-1}
\eqn\ertre
$$
where the superscript is the $\R^+$ charge. The corresponding values of $\rho$
for the three terms
on the right hand side of \ertre\ are $0, {1\over 2}$ and $-{1\over 2}$,
respectively.
If there are no metastable non-BPS states with higher or lower values of $\rho$
in the spectrum, then   for the coupling constant corresponding to the $\R^+$
factor in \mos,   the perturbative states
at strong coupling are $n$ {\it non-chiral }  strings (i.e. ordinary strings
satisfying no
self-duality condition) coupling to $G^+_a$ and transforming as an
{\bf n} of $SL(n,\R)$ while those at weak coupling are
$n$ {\it non-chiral } strings coupling to $G^-_a$ and transforming as an
${\bf n'}$ of $SL(n,\R)$.

The case $n=1$ is straightforward to interpret:
the decomposition
$${O(5,21)
\over O(5)\times O(21)} \to  {O(4,20)
\over O(4)\times O(20)}\times \R^+ \times \R^{24}
 \eqn\mosw $$
corresponds to the weak (or strong) coupling limit of the type IIB string
[\AM]; the $\R^+$ factor
corresponds to the type IIB  string coupling constant, given by the exponential
of the  dilaton
expectation value, while ${O(4,20)
\over O(4)\times O(20)}$ is the moduli space of conformal field theories on
$K_3$ [\AMK].
At weak coupling, the perturbative states correspond to the fundamental type
IIB string, while
at strong coupling we obtain the dual RR string, which is the fundamental
string of the dual theory
[\Oopen].

The decomposition
$${O(5,21)
\over O(5)\times O(21)} \to  {O(3,19)
\over O(3)\times O(19)}\times {SL(2,\R)\over U(1)}\times \R^+ \times \R^{45}
 \eqn\mosc $$
appears to correspond to expanding in the volume of the $K_3$ surface [\AM]:
${O(3,19)
\over O(3)\times O(19)}$ is the moduli space of fixed volume $K_3$ metrics,
${SL(2,R)\over U(1)}$ is the  moduli space of type IIB string theories in
$D=10$ and $\R^+ $ is the
   $K_3$ volume. However, the perturbative spectrum consists of {\it two}
strings, which is unlike
any conventional string theory.  (The only way out would be if there were some
metastable states dominating the perturbation theory, such as Kaluza-Klein
modes on $K_3$. Note however that no such states should occur in the previous
example of $n=1$ if the type IIB string interpretation is to hold.)

The  perturbative spectra arising for $n>1$ are remarkable in that they are
quite unlike
anything that has been seen in the cases considered
so far (again, unless some metastable states dominate). Recall that for a
supergravity theory compactified on an $n$-torus, the
perturbative spectrum for the coupling constant corresponding to the inverse
volume of the torus
consists of $n$ charged
$0$-brane multiplets that become massless supergravity multiplets in the large
volume limit and
which transform as an {\bf n} under the $SL(n)$ torus mapping class group.
Each of these $n$ multiplets carries
a minimal charge $e_0$ with respect to a corresponding gauge field and has
partners with charges
$me_0$ for all integers $m$, so that the spectrum includes $n$ Kaluza-Klein
towers. The same
perturbative spectrum emerged from superstring theories compactified on an
$n$-torus, when expanded
with respect to the inverse volume of the torus. Here we are finding something
rather different: $n$
superstrings instead of $n$ super-0-branes. Furthermore, each of these  has a
charge
$q_0$ with respect   to a particular 2-form
$G^+$ and has partners of   charge $mq_0$ for all integers $m$, so that we are
obtaining
Kaluza-Klein-like towers of {\it superstrings}.
 These facts, together with the
presence of the
${SL(n,\R)/O(n)}$ factor in
\mos\ and comparison with the various limits of the
non-chiral supergravities considered in the last section,
suggest that the theory could have an interpretation in terms of a
compactification of some
theory on an $n$-torus (or perhaps on some space that is locally $T^n$, such as
an orbifold of $T^n$).
However, the required theory  cannot be any known theory compactified on a
torus in the conventional
way from  $n+6$ dimensions, as the wrong spectrum would emerge. For example,
the case $n=5$ would
require reduction from 11 dimensions. Reducing the  11-dimensional M-theory on
$T^5$ gives
the usual type II string theory in $D=6$ which is rather different from what we
want, although it is
possible that reduction of M-theory on some  orbifold of
$T^5$ could   work here.

To get more insight, consider the compactification of the chiral theory to
5-dimensions on a circle
or radius $R$, as this gives the 5-dimensional heterotic theory [\Witten], with
moduli space
$${O(5,21)
\over O(5)\times O(21)}  \times \R^ +
 \eqn\fivmod $$
Consider the coupling constant $g$ such that the limit $g \to \infty$
corresponds to
the degeneration \mos.
Taking $g \to \infty$   corresponds to the decompactification to $5+n$
dimensions:- the
heterotic string in  5-dimensions arises from the heterotic string in $5+n$
dimensions on
compactification on a torus $T^n$ of volume $V$, and the coupling constant $g$
of the 5-dimensional
theory corresponds to the volume $V$, so that $g \to \infty$ corresponds to $V
\to \infty$. Thus
taking $g \to \infty$, we recover Lorentz invariance in $5+n$ dimensions. On
the other
hand, taking $R \to \infty$ for fixed $g$ we recover 6-dimensional Lorentz
invariance.
This suggests taking
both decompactification limits together, $g \to \infty, R \to \infty$ might
give  a theory a
$(6+n)$-dimensional theory, and in particular   an 11-dimensional theory for
$n=5$.

Unfortunately, the situation is not quite so simple as one has to be careful as
to how the limits are
taken.
Consider the case $n=2$ for example. The string coupling constant $g_7$ of the
$D=7$ heterotic string
is given by $g_7^2=g^3R^2$, so that on taking the limit $g \to \infty$ one must
take $R \to 0$
holding $g^3R^2$ constant if one is to obtain the heterotic string with finite
coupling $g_7$.
However, the heterotic string in $D=7$ is conjectured to be equivalent to
M-theory compactified on
a $K_3$ of volume $V=g_7^{4/3}$ [\Witten], so that   the limit $g_7 \to \infty$
corresponds to the
limit in which the volume of the $K_3$ becomes infinite, so that the theory
decompactifies and
11-dimensional Lorentz invariance is regained.
It is not clear whether this is related to  the limit we are interested in
here, which is given by
first taking
$R\to
\infty$ to regain the type IIB string compactified on $K_3$ and then taking
$g\to \infty$ so that
the moduli space decomposes as in \mosc. However, it is plausible that the
latter limit might define
a theory in more than 6 dimensions.

Thus the strong coupling limit corresponding to  \mos\ of the chiral $d=6$
theory, if it exists,
appears to be
 a theory  in at least six dimensions and   probably more, which has an
$n$-dimensional lattice
of superstrings in the perturbative spectrum, acted on by $SL(n,\Z)$. If, as
suggested above, it
arises from some theory compactified on
$T^n$ in the limit of large torus volume, then the limiting theory should live
in (at least)
$6+n$-dimensions  and  should have a moduli space
$${O(5-n,21-n)
\over O(5-n)\times O(21-n)}\eqn\modN$$
Note the absence of any dilaton-like $\R^+$ factor.
For $n=5$  there are two distinct limits \mos\ (cf [\AM]), so there could be
two distinct  theories
in (at least) 11 dimensions in this case, and these  would have no scalars and
so no coupling
constants and  no perturbation theory.
It will be convenient to refer to these theories as N-theories. There do not
appear to be any known
supergravity theories that could serve as the low-energy limits of these
N-theories, so if these
N-theories do exist, then either they do not have a low-energy effective field
theory, or there are
some new supergravity-type theories that are yet to be found.

So what could these N-theories be? One possibility is that they arise from
orbifold
compactifications of M-theory. Recall that the heterotic string in
$10-m$ dimensions can be obtained by compactifying M-theory on $T^m\times
S^1/\Z_2$ and the
5-dimensional heterotic string is equivalent to type IIB on $K_3\times S^1$ and
hence to M-theory on
$K_3\times T^2$. Moreover, $K_3$ can be obtained by blowing up the orbifold
singularities of
$T^4/\Z_2$. Then the N-theory corresponding to the degeneration \mos\ could
arise from M-theory
compactified on some orbifold of $T^{5-n}$, so that the N theory with $n=0$ --
the chiral $D=6$
theory -- would correspond to M-theory on some orbifold of $T^5$, and the
degeneration \mos\ would
correspond to a limit in the moduli space of the orbifold of $T^5=T^n\times
T^{5-n}$ in which the
volume of the $T^n$ became large. \foot{After this paper appeared, it was
argued by Dasgupta and Mukhi and by Witten that
the chiral theory in six dimensions given  by compactifying the type IIB string
on $K_3$ can also be obtained from M-theory by compactifying on the orbifold
$T^5/\Z_2$.}

Another intriguing possibility invokes a theory in twelve dimensions.
Suppose that the N-theory corresponding to \mos\ lives in $d=6+n$ dimensions.
When compactified on a circle, the N-theory in $d$ dimensions should give the
heterotic string in
$d-1$ dimensions.
Indeed, the modulus for the circle would give an extra $\R^+$ factor to \modN,
so that the correct
heterotic string moduli space emerges.
Consider for concreteness the $d=8$ N-theory which has moduli space
$$ {O(3,19)
\over O(3)\times O(19)}
\eqn\eimod$$
and which when compactified on $S^1$ should give the $d=7$ heterotic string,
which is
conjectured to  be equivalent to M-theory on $K_3$ [\Witten]. However, \eimod\
is the moduli space
for fixed-volume Ricci-flat metrics on $K_3$ and this suggests that $d=8$
N-theory might arise from
some theory in 12 dimensions compactified on $K_3$. Let us suppose that there
is such a  theory in
12 dimensions -- we shall refer to it as \lq Y-theory' -- and see what this
would imply. The
dynamics of Y-theory must be such that when compactified on $K_3$ the moduli
space is not the space
of all Ricci-flat metrics but only those of fixed volume, so that it is not a
conventional theory of
gravity.
This could come about, for example,  if
Y-theory had some conformal invariance which led to the volume being a gauge
degree of freedom,
or if
it was a  theory of gravity subject to certain constraints, which removed the
volume modulus.
  On further compactifying on $S^1$, one obtains the $D=7$ heterotic string,
corresponding
to M-theory  compactified on
$K_3$, and the  radius of the circle provides the extra degree of freedom
corresponding to the $K_3$
volume. This further suggests that Y-theory in $D=12$   can be compactified on
$S^1$ in such a way
that
\lq M-theory'   emerges in $D=11$. Thus while compactifying from $D=11$ to
$D=10$, the radius of the $S^1$ gives rise to the dilaton in $D=10$, the
compactification from
$D=12$ to $D=11$ gives rise to the extra gravitational degree of freedom   that
is needed to give
unconstrained gravity in $D=11$.

This can be extended to the other degenerations. The degeneration \mos\ with $n
\ge 2$ would
correspond to Y-theory compactified on $\Xi^{6-n}$ and for $n\le 2$ perhaps to
Y-theory on
$K_3\times T^{2-n}$. Then a consistent picture  seems to emerge which is
similar to that
proposed for   the non-chiral $D=6$ theory and its limits in the previous
section.
It is conceivable that both
  the Y-theory picture and the orbifold picture
are correct and that there is a relation between them, which would be similar
to
  the relation between the Aspinwall-Morrison picture [\AM] and the
Horava-Witten
description [\Mwit]
of the theories described in the last section. The type IIB theory might also
be obtained from 12
dimensions, which would afford a geometrical interpretation of the $SL(2)$
symmetry, without
needing to go through 9-dimensional theories, as in [\bergort,\asptr]. Needless
to say, this is
rather speculative, but if M-theory, why not Y-theory?

%%%%%%%%%%%%%%%%%%%%%%%%%%%%%%%%
%%%%%%%%%%%%%%%%%%%%%%%%%%%%%%%%


\Appendix {A}

The rescaling of the metric
$$ \tilde g _{\mu \nu} = e^{2 \Lambda} \gmn
\eqn\git$$
leads to the following changes in the curvature
$$ \tilde R = e^{-2 \Lambda} \left[ R-2(D-1) \nabla ^2 \Lambda - (D-1)(D-2)
(\nabla \Lambda)^2
\right]
\eqn\rti$$

$$ \sqrt
 {\tilde g} \tilde R = e^{(D-2) \Lambda} \sqrt
 g \left[ R+ (D-1)(D-2) (\nabla \Lambda)^2\right] - 2 \sqrt
 g (D-1) \nabla^\mu ( e^{(D-2) \Lambda} \nabla _\mu \Lambda)
\eqn\grti$$
For an $n$-form potential $A_n = A_{ \mu _1 \mu _2 ... \mu _n } dx^{\mu_
1}dx^{\mu_ 2}
 \dots dx^{\mu_ n}$, the rescaling of the metric leads to
$$ \sqrt
 { g}  { \vert  dA_n  \vert ^2} \to \sqrt
 g e^{[D- 2(n+1)]\Lambda }  \vert dA_n \vert ^2
\eqn\dan$$

%%%%%%%%%%%%%%%%%%%%%%%%%%%%%%%Acknowledgements%%%%%%%%%%%%%%%%%%%%%%%%%%
%%%%%%
%\vskip 0.5cm
\noindent{\bf Acknowledgements}: I would like to thank Michael Green, Ashoke
Sen and Paul Townsend for valuable
discussions.
%%%%%%%%%%%%%%%%%%%%%%%%%%%%%%%%%%%%%%%%%%%%%%%%%%%%%%%%%%%%%%%%%%%%%%%%%
%%%%%%

\refout
\bye

\end